\documentclass[aps,prd,tightenlines,showpacs,amsmath,amssymb]{revtex4}%
\usepackage{graphicx}% Include figure files
\usepackage{epsfig}
\usepackage{graphics,color}
\usepackage{amssymb}
\usepackage{hyperref}
\usepackage{amsbsy}
\usepackage{bm}
\newcommand{\be}{\begin{equation}}
\newcommand{\ee}{\end{equation}}
\newcommand{\bse}{\begin{subequations}}
\newcommand{\ese}{\end{subequations}}
\newcommand{\bea}{\begin{eqnarray}}
\newcommand{\eea}{\end{eqnarray}}

\newcommand{\bean}{\begin{eqnarray*}}
\newcommand{\eean}{\end{eqnarray*}}

\begin{document}
\title{Nontopological Soliton in the Polyakov Quark Meson Model}
\author{Jinshuang Jin}
\author{Hong Mao}
\email {mao@hznu.edu.cn (corresponding author)}
\address{Department of Physics, Hangzhou Normal University, Hangzhou 310036, China }

%\date{}

\begin{abstract}
Within a mean field approximation, we study a nontopological soliton solution of the Polyakov quark-meson model in the presence of a fermionic vacuum term with two flavors at
finite temperature and density. The profile of the effective potential exhibits a stable soliton solution below a critical temperature $T\leq T_{\chi}^c$ for both the
crossover and the first-order phase transitions, and these solutions are calculated here with appropriate boundary conditions. However, it is found that only if $T\leq T^c_d$,
the energy of the soliton $M_N$ is less than the energy of the three free constituent quarks $3M_q$. As $T> T^c_d$, there is an instant delocalization phase transition from
hadron matter to quark matter. The phase diagram together with the location of a critical end point (CEP) has been obtained in $T$ and $\mu$ plane. We notice that two critical
temperatures always satisfy $T^c_d\leq T_{\chi}^c$. Finally, we present and compare the result of thermodynamic pressure at zero chemical potential with lattice data.
\end{abstract}


\maketitle

\section{Introduction}
It is widely believed that at sufficiently high temperatures and densities there is a quantum chromodynamics (QCD) phase transition between normal nuclear matter and
quark-gluon plasma (QGP), where quarks and gluons are no longer confined in hadrons\cite{Rischke:2003mt}\cite{Yagi:2005yb}. The study of the QCD phase transition is
experimentally supported by the heavy-ion collisions at ultrarelativistic energies, accomplished in the most sophisticated accelerators, such as the Relativistic Heavy-Ion
Collider (RHIC) at Brookhaven National Laboratory and the Large Hadron Collider (LHC) at CERN. These conducted experiments provide us with the opportunity to infer fundamental
information about which type phase of matter, hadronic or quark-gluon plasma, is stabilized in the various regimes. In order to explore a wider range of the QCD phase
transition up to several times the normal nuclear matter density, the new Facility for Antiproton and Ion Research (FAIR) at Darmstadt, the Nuclotron-based Ion Collider
Facility (NICA) at the Joint Institute for Nuclear Research (JINR) in Dubna and the Japan Proton Accelerator Research Complex (J-PARC) at JAEA and KEK, will make such extreme
conditions possible through collisions\cite{Fukushima:2010bq}. Therefore, it will be possible to test the theoretical predictions about the hadron-quark phase transition at
high density but moderate temperature.

On the theoretical side, the property of confinement which becomes relevant at large distances or equivalently low energies, has hindered the development of analytical and
numerical methods capable of describing the low-energy nonperturbative cases, especially if baryons are involved. Therefore, the challenge for nuclear physicists remains to
find models that can bridge the gap between the fundamental theory and our wealth of knowledge about low energy phenomenology. Moreover, these models should be successful in
explaining empirical facts at low energies, for example, the dynamical breaking of chiral symmetry and the confinement, which are both intimately related to the
nonperturbative structure of the QCD vacuum. To mention a few, these effective models are the MIT bag model\cite{Chodos:1974je}, the Nambu-Jona-Lasinio
(NJL)\cite{Nambu:1961tp}\cite{Vogl:1991qt} and the linear sigma model\cite{GellMann:1960np} for quark matter, while others are the Brueckner-Hartree-Fock (BHF)
theory\cite{Ring:1980} and the relativistic mean-field (RMF) models\cite{Serot:1986} for nuclear matter.

Since the strongly interacting matter at very high energy should have quarks and gluons as the degrees of freedom, while nucleons and mesons are the relevant degrees of
freedom in the hadron phase. On one hand, the phenomenology of the hadron-quark phase transition is often studied with the above mentioned effective
models\cite{Yasutake:2012dw,Lourenco:2012dx,Shao:2011fk}, where the BHF theory including the realistic baryon baryon interaction or the RMF models are used to describe hadron
phase, while on the other, the quark phase is treated as a thermodynamic bag model or simulated by the NJL model (or its modernized version, the Polyakov Nambu-Jona-Lasinio
model)\cite{Costa:2010zw}. The motivation for these studies, aims at rendering two models with different degrees of freedom compatible. In more detail, the MIT bag model and
PNJL models do describe the very well known properties of quark matter, but they fail to reproduce the bulk nuclear matter and finite nuclei properties. On the other side, the
BHF theory and the RMF models which are constructed for nuclear matter are often questionable when extending to investigate high density regimes as it is common for neutron
stars. Then the approach of considering the mixed phase of hadron and quark matter becomes important, and the proper equation of state (EOS) for the hadron-quark phase
transition could be derived based on the Gibbs conditions for phase equilibrium. Unfortunately, this kind of study is applicable only if the QCD phase transition is of first
order. According to a recent study on QCD phase diagram based on chiral effective models, including quark and meson fluctuations via the functional renormalization
group\cite{Herbst:2013ail}, the biggest part of the QCD phase diagram shows crossover transitions rather true first-order ones. Thus, it is necessary to search for an
alternative effective model which can provide a proper description of hadron-quark phase transition beyond the first-order transition cases, including the crossover
situations, while keeping at the same time the correct degrees of freedom in quark phase and hadron phase intact. The nontopological soliton model rooted on the Polyakov Quark
Meson Model\cite{Schaefer:2007pw} appears to fulfill the requirement.

The Polyakov Quark Meson Model has so far facilitated the investigation of the full QCD thermodynamics and phase structure at zero and finite quark chemical potential, while
it has been shown that the bulk thermodynamic predictions of the model agree well with the Lattice QCD
data\cite{Herbst:2013ail,Schaefer:2007pw,Marko:2010cd,Herbst:2010rf,Gupta:2011ez,Skokov:2010sf,Tiwari:2012yy,Mao:2009aq,Gupta:2009fg,Schaefer:2009ui}. On the other hand,
starting from the same Lagrangian, bound states (solitons) of valence quarks can be constructed through the interaction with $\sigma$ and $\pi$
mesons\cite{Birse:1983gm,Kahana:1984dx}. Such a nontopological soliton give rise to nucleons in the hadron phase. Moreover, the nontopological soliton model has been proven to
be a successful approach for the description of the static properties of nucleons in
vacuum\cite{Birse:1983gm,Alberto:1988xj,Bernard:1988db,Alberto:1990ru,Goeke:1988hp,Aly:1998wg}. Combining these two features together, while also requiring a soliton embedded
in a thermal medium, the model provides a suitable working scheme to simultaneously study both the restoration of chiral symmetry and the possible dissolution of the soliton,
which simulates the deconfinement transition of nuclear matter to quark matter.

In fact, the nucleon has been previously investigated in Ref\cite{Christov:1991ry} by employing the chiral soliton model and viewing it as a $B=1$ chiral soliton in a
cold quark medium. However, the parameters $f_{\pi}$, $m_{\pi}$ and $m_{\sigma}$ were chosen to be the medium-modified meson values within the NJL model. For finite
temperature, Abu-Shady and Mansour have studied nucleon properties \cite{AbuShady:2012zza} by employing the one-loop phenomenological mesonic potential\cite{Hong1997clj} and
the coherent-pair approximation\cite{Goeke:1988hp}\cite{Aly:1998wg}. Furthermore, the nucleon properties as well as the thermodynamics of the system both at finite temperature
and density are examined in Refs \cite{Mao:2013qu,Zhang:2015vva,Mansour:2015yha}. However, these studies based on the chiral soliton model or other nontopological soliton
models\cite{Reinhardt:1985nq,Gao:1992zd,Li:1987wb,Mao:2006zp} suffer from two problems: the one is that they only predict a first-order phase transition and the other is
that the critical temperature is extremely low ($T_c\sim 110 \mathrm{MeV}$) as compared with lattice data. In this work, we will improve theses previous studies by combing the
chiral soliton model with the Polyakov loop field. Such an extension will allow us to inspect both the crossover and first-order QCD phase transitions and compare directly
with the lattice QCD simulations.

The structure of the paper is as follows: in the next section we introduce the PQM model with two quark flavors. In Sect. III, after obtaining the effective potential in the
mean field approximation, we explore the possible stable soliton solutions in the model. Section IV is devoted to derive the equations of motion of the nontopological soliton
model both in vacuum and at finite temperature and density. Section V contains the static properties of nucleon at finite temperature and density and the phase diagram at
$T-\mu$ plane. The study of the hadron-quark phase transition is presented in section VI. We conclude with a summary and discussions in Sec.VII.

\section{The Model}
We will be working in a generalized Lagrangian of the quark-meson model for $N_f=2$ quarks and $N_c=3$ color degrees with quarks coupled to a spatially-constant
time-dependent background gauge field representing Polyakov loop dynamics (the Polyakov-quark-meson model or the PQM in short). The Lagrangian reads\cite{Schaefer:2007pw}
\begin{equation}
{\cal L}=\overline{\psi} \left[ i\gamma ^{\mu} D _{\mu}-
g(\sigma +i\gamma _{5}\vec{\tau} \cdot \vec{\pi} )\right] \psi
+ \frac{1}{2} \left(\partial _{\mu}\sigma \partial ^{\mu}\sigma +
\partial _{\mu}\vec{\pi} \cdot \partial ^{\mu}\vec{\pi}\right)
-U(\sigma ,\vec{\pi})-\mathbf{\mathcal{U}}(\Phi,\Phi^*,T) .\\
\label{Lagrangian}
\end{equation}
Here, we have introduced a flavor-blind Yukawa interaction of strength $g$, coupling the isodoublet spin-$\frac{1}{2}$ quark fields $\psi=(u,d)$, with the
spin-$0$ isosinglet $\sigma$ and the isotriplet pion field $\vec{\pi} =(\pi _{1},\pi _{2},\pi _{3})$. In addition, there exists a spatially homogeneous time-dependent gauge
field represented by the Polyakov loop potential. The coupling of the quarks with the uniform temporal background gauge field is implemented through the covariant derivative
$D_{\mu}=\partial_{\mu}-i A_{\mu}$ and the spatial components of the gauge fields have vanishing background $A_{\mu}=\delta_{\mu 0}A_0$.

The purely mesonic potential for the $\sigma$ and $\vec{\pi}$ is defined as
\begin{equation}
U(\sigma ,\vec{\pi})=\frac{\lambda}{4} \left(\sigma ^{2}+\vec{\pi} ^{2}
-{\vartheta}^{2}\right)^{2}-H\sigma -\frac{m^{4}_{\pi}}{4 \lambda}+f^{2}_{\pi}m^{2}_{\pi},
\label{mpot}
\end{equation}
and the minimum energy occurs for chiral fields $\sigma$ and $\vec{\pi}$ restricted to the chiral circle in the physical vacuum:
\begin{eqnarray}
\sigma^2+\vec{\pi}^2=f_{\pi}^2,
\label{ccircle}
\end{eqnarray}
where $f_{\pi}=93 \mathrm{MeV}$, corresponds to the pion decay constant and $m_{\pi}=138 \mathrm{MeV}$ is the pion mass. The last two constant terms in Eq.(\ref{mpot}) are
used to guarantee that the energy of the vacuum in the absence of quarks is zero. The constant $H$ is fixed by the PCAC relation which gives $H=f_{\pi}m_{\pi}^{2}$.

The quantity $\mathbf{\mathcal{U}}(\Phi,\Phi^*,T)$ is the Polyakov-loop effective potential. The Polyakov loop field $\Phi$ is defined as the thermal expectation value of
the color-trace of the Wilson loop along the temporal direction
\begin{eqnarray}
\Phi=(\mathrm{Tr}_c L)/N_c, \qquad \Phi^*=(\mathrm{Tr}_c
L^{\dag})/N_c.
\end{eqnarray}
The Polyakov loop $L$ is a matrix in color space and explicitly
given by
\begin{eqnarray}
L(\vec{x})=\mathcal{P}\mathrm{exp}\left[i\int_0^{\beta}d \tau
A_4(\vec{x},\tau)\right],
\end{eqnarray}
with $\beta=1/T$ being the inverse of temperature and $A_4=iA^0$. In the so-called Polyakov gauge, the Polyakov-loop matrix can be given as a diagonal representation
\cite{Fukushima:2003fw}. Within this diagonal representation, $\Phi$ and $\Phi^*$ are complex scalar fields. Their mean values are related to the free energy of a static,
infinitely heavy test quark (anti-quark) at spatial position $\vec{x}$. The Polyakov loop expectation value $\langle \Phi \rangle$ vanishes in the confined phase where the
free energy of a single heavy quark diverges, while in the deconfined phase it takes a finite value since the center symmetry becomes spontaneously broken
\cite{Polyakov:1978vu}.

The temperature dependent effective potential $\mathbf{\mathcal{U}}(\Phi,\Phi^*,T)$ is constructed to reproduce the thermodynamical behavior of the Polyakov loop for the pure
gauge case in accordance with lattice QCD data, and it has the $Z(3)$ center symmetry like the pure gauge QCD Lagrangian. In the absence of quarks, we have $\Phi=\Phi^*$ and
the Polyakov loop is taken as an order parameter for deconfinement. For low temperatures, $\mathbf{\mathcal{U}}$ has a single minimum at $\Phi=0$, while at high temperatures
it develops a second one which turns into the absolute minimum above a critical temperature $T_0$, and the $Z(3)$ center symmetry is spontaneously broken. The simplest $Z(3)$
symmetric polynomial form based on a Ginzburg-Landau ansatz is proposed in Ref.\cite{Ratti:2005jh}
\begin{eqnarray}
\frac{\mathbf{\mathcal{U}}(\Phi,\Phi^*,T)}{T^4}=-\frac{b_2(T)}{4}(|\Phi|^2+|\Phi^*|^2)-\frac{b_3
}{6}(\Phi^3+\Phi^{*3})+\frac{b_4}{16}(|\Phi|^2+|\Phi^*|^2)^2,
\end{eqnarray}
with
\begin{eqnarray}
b_2(T)=a_0+a_1\left(\frac{T_0}{T}\right)+a_2\left(\frac{T_0}{T}\right)^2+a_3\left(\frac{T_0}{T}\right)^3.
\end{eqnarray}
A precise fit of the constants $a_i,b_i$ is performed to reproduce the lattice data for the pure gauge theory thermodynamics, as also the behavior of the Polyakov loop as a
function of temperature. The corresponding parameters are
\begin{eqnarray}
a_0=6.75,\qquad a_1=-1.95,\qquad a_2=2.625, \nonumber \\
 a_3=-7.44,\qquad
b_3=0.75,\qquad b_4=7.5.
\end{eqnarray}
Originally, the critical temperature $T_0$ for deconfinement in the pure gauge sector is fixed at $270$ MeV, in agreement with the lattice results. However, in fully dynamical
QCD, fermionic contributions and the matter backreaction modify the pure gauge potential to an effective glue potential, which carries a flavor and chemical potential
dependence of $T_0$. The actual value of $T_0$ for two quark flavors is $T_0=208 \mathrm{MeV}$\cite{Schaefer:2007pw}\cite{Gupta:2011ez}.

A convenient framework of studying phase transitions is the thermal field theory. Within this framework, the finite temperature effective potential is an important and useful
theoretical tool. In this section, in order to investigate the temperature and the chemical potential dependence of the nontopological soliton, let us consider a spatially
uniform system in thermodynamical equilibrium at temperature $T$ and quark chemical potential $\mu$. In general, the grand partition function reads
\begin{eqnarray}
\mathcal{Z}&=& \mathrm{Tr exp}[-(\hat{\mathcal{H}}-
\mu \hat{\mathcal{N}})/T] \nonumber \\
&=& \int\prod_a \mathcal{D} \sigma \mathcal{D} \pi_a \int
\mathcal{D}\psi \mathcal{D} \bar{\psi} \mathrm{exp} \left[ \int_x
(\mathcal{L}+\mu \bar{\psi} \gamma^0 \psi )
\right],
\end{eqnarray}
where $\int_x\equiv i \int^{1/T}_0 dt \int_V d^3x$, $V$ is the volume of the system and $\mu=\mu_B /3 $ for the homogeneous background field.

We evaluate the partition function in the mean-field approximation similar to the work of \cite{Scavenius:2000qd}. Thus we replace the meson fields by their expectation values
in the action. In other words, we neglect both quantum and thermal fluctuations of the meson fields. The quarks and antiquarks are retained as quantum fields. The integration
over the fermions yields a determinant which can be calculated by standard methods\cite{Kapusta:2006pm}. This generates an effective potential for the mesons. Finally, we
obtain the thermodynamical potential density as
\begin{eqnarray}\label{potential}
\Omega(T,\mu)=\frac{-T \mathrm{ln}
\mathcal{Z}}{V}=U(\sigma ,\vec{\pi} )+\mathbf{\mathcal{U}}(\Phi,\Phi^*,T)+\Omega_{\bar{\psi}
\psi} ,
\end{eqnarray}
with the quarks and antiquarks contribution
\begin{eqnarray}\label{omegapsi}
\Omega_{\bar{\psi} \psi} =\Omega_{\bar{\psi} \psi} ^{\mathrm{v}}+\Omega_{\bar{\psi} \psi}^{\mathrm{th}} =-2N_f N_c\int \frac{d^3\vec{p}}{(2
\pi)^3}E_q  -2 N_f T \int \frac{d^3\vec{p}}{(2
\pi)^3} \left[ \mathrm{ln} g_q^+ + \mathrm{ln} g_q^- \right],
\end{eqnarray}
where, $N_f=2$, $N_c=3$ and $E_q=\sqrt{\vec{p}^2+M_q^2}$ is the valence quark and antiquark energy for $u$ and $d$ quarks, and the constituent quark (antiquark) mass $M_q$ is
defined as $M_q=g \sigma_v$ together with $\sigma_v\equiv\sqrt{\sigma^2+\vec{\pi}^2 }$. The first term of Eq.(\ref{omegapsi}) denotes the fermion vacuum one-loop contribution,
regularized by the ultraviolet cutoff. In the second term $g_q^+$ and $g_q^-$ are defined as taking trace over color space
\begin{eqnarray}
 g_q^+ &=& \left[ 1+3(\Phi+\Phi^*e^{-(E_q-\mu)/T})\times e^{-(E_q-\mu)/T}+e^{-3(E_q-\mu)/T} \right],  \\
 g_q^- &=& \left[ 1+3(\Phi^*+\Phi e^{-(E_q+\mu)/T})\times e^{-(E_q+\mu)/T}+e^{-3(E_q+\mu)/T} \right].
\end{eqnarray}

The fermion vacuum one-loop contribution $\Omega_{\bar{\psi} \psi} ^{\mathrm{v}}$ is frequently omitted. In this work, we shall include the effect of vacuum fluctuation on the
thermodynamics. This term can be properly renormalized by using the dimensional regularization scheme as done for the two-flavor case in
Refs.\cite{Gupta:2011ez,Skokov:2010sf,Tiwari:2012yy}, and the renormalized contribution of the fermion vacuum loop reads
\begin{eqnarray}\label{omegareg}
\Omega_{\bar{\psi} \psi} ^{\mathrm{v}}=\Omega_{\bar{\psi} \psi} ^{\mathrm{reg}}=-\frac{N_c N_f}{8\pi^2} M_q^4 \mathrm{ln}(\frac{M_q}{\Lambda}),
\end{eqnarray}
where $\Lambda$ is the arbitrary renormalization scale. It is worth to note that the thermodynamic potential and all physical observable are independent of the choice of
$\Lambda$, and the $\Lambda$ dependence can be neatly cancelled out by redefining the parameters in the model.

Now the first term in the right hand side of Eq.(\ref{omegapsi}), describing the vacuum contribution, will be replaced by the appropriately renormalized fermion vacuum
contribution as given in Eq.(\ref{omegareg}). Accordingly, the thermodynamic grand potential in the presence of appropriately renormalized fermionic vacuum contribution in the
Polyakov quark meson model will be written as
\begin{eqnarray}\label{omegamf}
\Omega_{\mathrm{MF}}(T,\mu,\sigma_v,\Phi,\Phi^*) =\Omega_{\mathrm{M}}(\sigma_v)+\mathbf{\mathcal{U}}(\Phi,\Phi^*,T)+\Omega_{\bar{\psi} \psi}^{\mathrm{th}} ,
\end{eqnarray}
Here, for convenience we define a new mesonic potential
\begin{eqnarray}
\Omega_{\mathrm{M}}(\sigma_v)=U(\sigma ,\vec{\pi} )+\Omega_{\bar{\psi} \psi} ^{\mathrm{reg}},
\end{eqnarray}
which is independent of the temperature $T$ and the chemical potential $\mu$. Minimizing the thermodynamical potential in Eq.(\ref{omegamf}) with respective to $\sigma_v$,
$\Phi$ and $\Phi^*$, we
obtain a set of equations of motion
\begin{eqnarray}
\frac{\partial \Omega_{\mathrm{MF}}}{\partial \sigma_v}=0, \qquad \frac{\partial \Omega_{\mathrm{MF}}}{\partial
\Phi}=0, \qquad \frac{\partial \Omega_{\mathrm{MF}}}{\partial \Phi^*}=0.
\end{eqnarray}
The set of equations can be solved for the fields as functions of temperature $T$ and chemical potential $\mu$, and the solutions of these coupled equations determine the
behavior of the chiral order parameter $\bar{\sigma}_v$ and the Polyakov loop expectation values $\bar{\Phi}$, $\bar{\Phi}^*$ as a function of $T$ and $\mu$.

There are two values of the constants left in the model which we need to fix, namely, $m_{\sigma}$ and $g$. Unlike the pion meson, the mass of the sigma meson still has
a poorly known value, but the most recent result of the Particle Data Group considers that $m_{\sigma}$ can vary from $400$ MeV to $550$ MeV with full width
$400-700$ MeV\cite{Agashe:2014kda}. The coupling constant $g$ is usually fixed by the constituent quark mass in vacuum within the range of $300\sim 500$ MeV, which gives $g
\simeq 3.3\sim 5.3$. In this work we take $m_{\sigma}=472$ MeV and $g=4.5$ as the typical values. It has been proved in Ref.\cite{Mao:2013qu} that this set of parameters can
describe the properties of nucleon in vacuum successfully.

\begin{figure}
\includegraphics[scale=0.36]{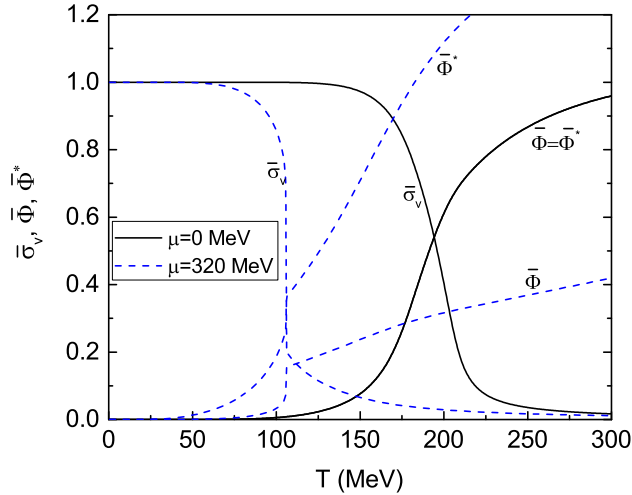}
\caption{\label{Fig01} (Color online) The normalized chiral order parameter $\bar{\sigma}_v$ and the Polyakov loop expectation values $\bar{\Phi}$, $\bar{\Phi}^*$ as functions of
temperature for $\mu = 0 \mathrm{MeV}$ and $\mu = 320 \mathrm{MeV}$. The solid curves are for  $\mu = 0 \mathrm{MeV}$ and the dashed curves are for $\mu = 320 \mathrm{MeV}$.}
\end{figure}

In figure \ref{Fig01}, the temperature dependence of the normalized chiral order parameter $\bar{\sigma}_v$ and the Polyakov loop expectation values $\bar{\Phi}$,
$\bar{\Phi}^*$ at $\mu = 0 \mathrm{MeV}$ and $\mu = 320 \mathrm{MeV}$ are shown in relative units. The temperature behaviors of the chiral condensate and Polyakov loop
condensate show that the system experiences a smooth crossover transition at zero chemical potential, while there is a first-order phase transition for larger chemical
potential because both the chiral order parameter and the Polyakov loop expectation values make jumps across the gap of the condensates near the critical temperature.
Traditionally, the temperature derivative of the chiral condensate  $\bar{\sigma}_v$ for $u$ and $d$ quarks has a peak at some specific temperature, which is established as
the critical temperature for the chiral phase transition for both the crossover and the first-order transitions. Thus, for zero chemical potential, the chiral restoration
occurs at $T_{\chi}^c \simeq 201  \mathrm{MeV}$, whereas for a relatively larger chemical potential $\mu = 320 \mathrm{MeV}$, the critical temperature moves to the lower
temperature region around $T_{\chi}^c \simeq 105  \mathrm{MeV}$.

Although different from the chiral phase transition, we are still not in a position to conclusively identify the deconfinement phase transition through the Polyakov loop
expectation values $\bar{\Phi}$, $\bar{\Phi}^*$ or their temperature derivatives \cite{Mao:2009aq}\cite{Fukushima:2008wg}. The temperature derivatives $\bar{\Phi}'$ and
${\bar{\Phi}^*}{'} $ do show one peak or more peaks for zero chemical potential or finite chemical potentials in calculations, but unfortunately these peaks are fake signals
for defining the critical temperature of the deconfinement. In the next section, based on the effective potential at finite temperature and finite chemical potential, we will
explain that there is no obvious clue to define the critical temperature of the deconfinement phase transition simply by using the Polyakov loop expectation values
$\bar{\Phi}$, $\bar{\Phi}^*$ or their temperature derivatives if $T<T_0$, even in the first-order transition region. This is a serious problem which appeared
already in the Polyakov quark meson model\cite{Schaefer:2007pw}\cite{Mao:2009aq} or the Polyakov-Nambu-Jona-Lasinio model\cite{Fukushima:2003fw}\cite{Fukushima:2008wg}, and
still persists in more recent theories. In the following, based on the nontopological soliton model, we will provide a distinct clarification on this point which
will allow us to propose a convincing definition of the deconfinement critical temperature.

\section{Effective potential and nontopological soliton}
The basic ideas behind the nontopological soliton are best illustrated by considering the original model: the Friedberg-Lee model\cite{Friedberg:1976eg} or its descendant
models\cite{Birse:1983gm,Kahana:1984dx,Birse:1991cx}. In these models, the confinement of quarks is approximated through their interaction with the phenomenological scalar
field, $\sigma$, which is introduced to describe the complicated nonperturbative features of the QCD vacuum. In mean field approximation, the $\sigma$ field has a bag like
soliton solution, named as nontopological soliton. This in contrast to topological defects which are stabilized by the topological properties of the vacuum manifold. The
existence of this kind of solution is closely related to a potential describing the nonlinear self-interactions of the $\sigma$ field. In general, the potential leading to the
soliton solution has three extrema: one local minimum corresponding to a perturbative vacuum state located at $\sigma  \simeq 0$, one absolute minimum corresponding to a
physical vacuum at its vacuum value $\bar{\sigma}_v$, and a local maximum lying between $0$ and $\bar{\sigma}_v$. Therefore, the soliton solution has a spherical cavity-like
structure: at large radius $r$, the $\sigma$ field assumes its vacuum value $\sigma_v$, but at small $r$, the $\sigma$ field has a value close to the second minimum of the
potential near zero. In the Friedberg-Lee model the quarks interact with a mean $\sigma$ field only, this means that, in the physical vacuum state the quark mass is more
than $1 \mathrm{GeV}$ which makes it energetically unfavorable for the quark to exist freely, so that the effective heavy quarks have to be confined in hadron bags. Sometimes
it is also called as ``absolute'' confinement, similar to the MIT bag model. However, as it is known for the chiral soliton model, pions can produce strongly attractive forces
among the quarks. Including mean pion fields also allows the meson fields to remain close to the minimum of the mexican hat potential, and so quarks possess
physical constituent masses in the physical vacuum. The state is to be bound if only the total energy of system is lower than the energy of three free constituent quarks in
the system, making thus transparent for considering the chiral soliton as a bound state in this work.

Normally, the self-interaction potential of the $\sigma$ field is chosen to have a quartic form in the nontopological soliton model, and the coefficients in the quartic
potential can be chosen so that they belong to the three typical forms as described as in a seminal work done by Goldflam and Wilets\cite{Goldflam:1981tg}. In their work, they
have shown that in order to ensure the stability of the two vacuum states and guarantee the existence of the stable soliton solution, it is indispensable for the potential of
the $\sigma$ field to exhibit three distinct extrema. In the following discussion, this will be also considered as a key criterion for determining whether there exist stable
soliton solutions for the mesonic fields and the Polyakov loop fields. Then with employing the thermodynamic grand potential in the presence of appropriately renormalized
fermionic vacuum contribution in Eq.(\ref{omegamf}), we can explore the possible nontopological soliton solutions owned by the model under this criterion.

However, in contrast to the chiral soliton and the Friedberg-Lee models, in the present study there are three order parameter variables $\sigma_v$, $\Phi$ and $\Phi^*$
in the grand canonical potential $\Omega_{\mathrm{MF}}$ in Eq.(\ref{omegamf}), so it is extremely difficult to investigate and demonstrate the effective potential at finite
temperature and chemical potential via evolving these variables simultaneously in such a large parameter space. In order to simplify the problem and provide a more
intuitive insight into the physics, we separate the study into two cases: (1) the mesonic field direction, and (2) the Polyakov loop field direction. For the first case, we
treat the $\sigma$ field as a variable in the grand canonical potential $\Omega_{\mathrm{MF}}$ while fixing the Polyakov loop fields on their expectation values $\bar{\Phi}$
nd $\bar{\Phi}^*$. In contrast, for the second case, the Polyakov loop $\Phi$ and $\Phi^*$ are consider as variables while the $\sigma$ field will maintain its expectation
value $\bar{\sigma}_v$ all the time.

\begin{figure}[thbp]
\epsfxsize=9.0 cm \epsfysize=6.5cm
\epsfbox{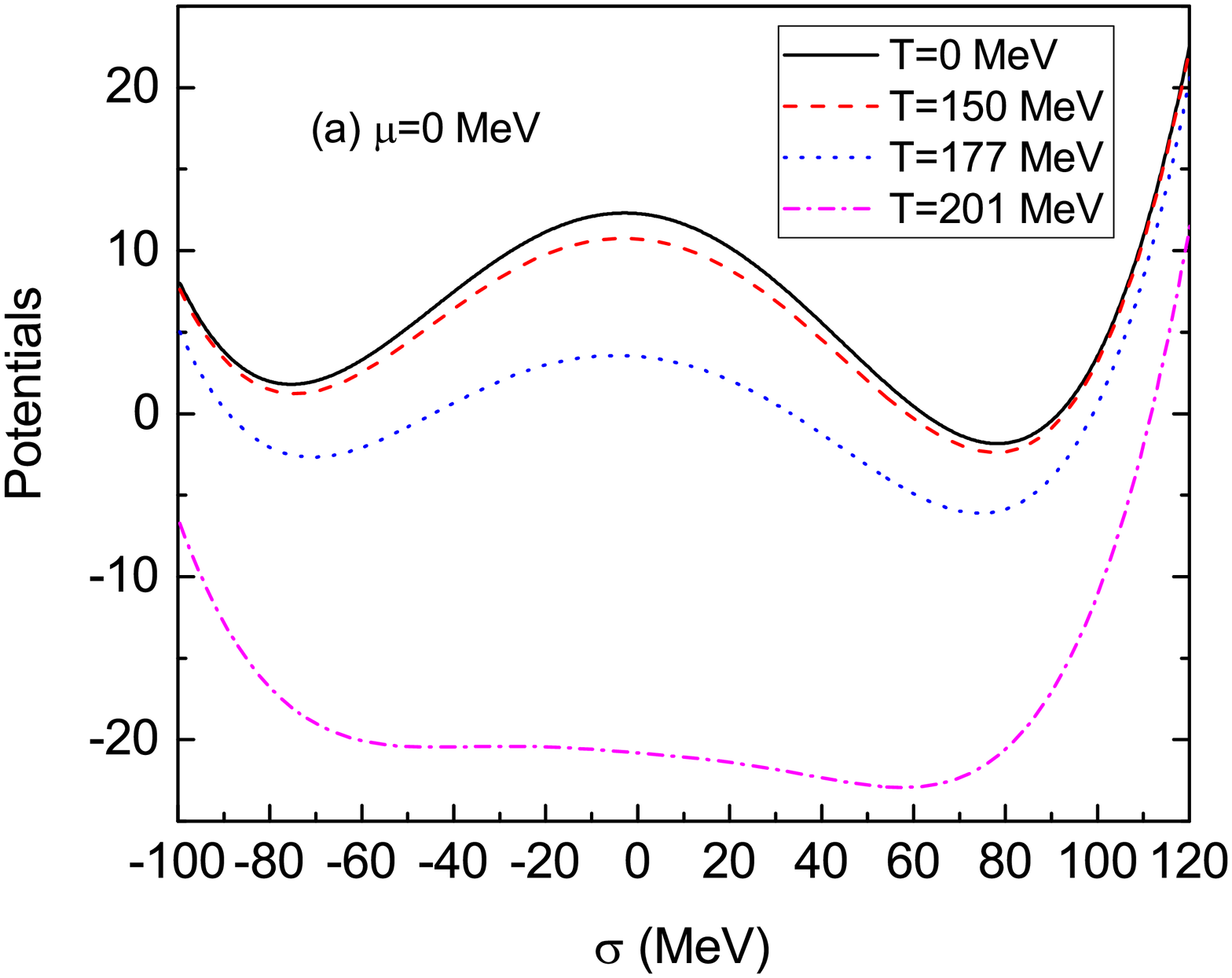}\hspace*{0.1cm} \epsfxsize=9.0 cm
\epsfysize=6.5cm \epsfbox{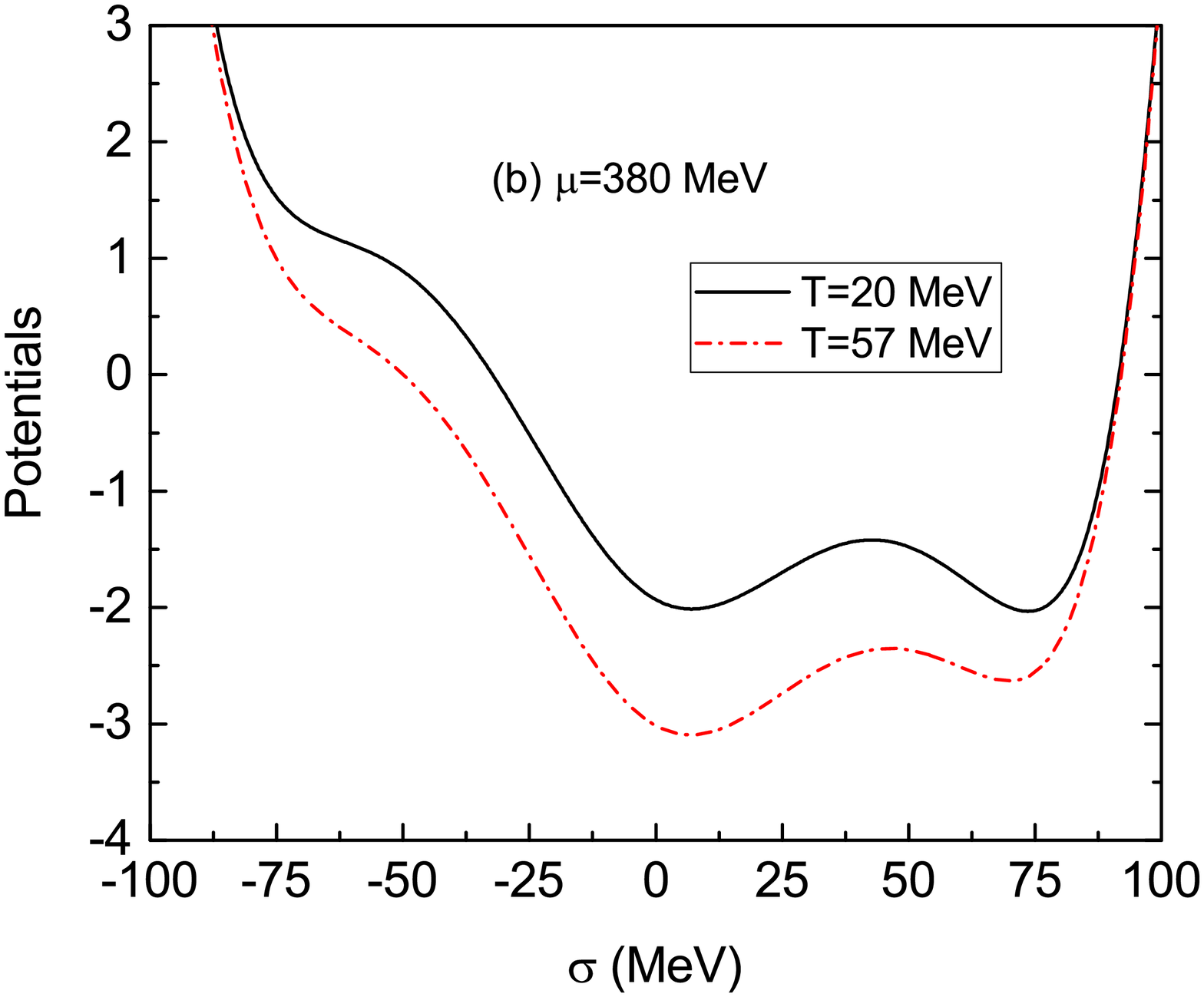}
%\vskip -0.05cm \hskip 0.15 cm \textbf{( a ) } \hskip 6.5 cm \textbf{( b )} \\
 \caption{(Color online) (a) The grand canonical potentials $\Omega_{\mathrm{MF}}$ as a function of the chiral order parameter $\sigma$ for $\mu=0 \mathrm{MeV}$ by fixing the Polyakov loop
on their expectation values. (b)The grand canonical potentials $\Omega_{\mathrm{MF}}$ as a function of the chiral order parameter $\sigma$ for $\mu=380 \mathrm{MeV}$ by fixing
the Polyakov loop on their expectation values. $\Omega_{\mathrm{MF}}$ is scaled by a factor of $f_{\pi}^4$.}
\label{Fig02}
\end{figure}

The first case is shown in Fig.\ref{Fig02}, where the left panel is for zero chemical potential and the right panel for $\mu=380 \mathrm{MeV}$. Here, the expectation value of
the pion field is chosen in the standard way as $\langle \vec{\pi} \rangle=0$. For $\mu=0 \mathrm{MeV}$, one clearly observes a smooth crossover of the symmetry breaking
pattern. The energy difference $\Delta \varepsilon$ between the global minimum and the local maximum of the potential decreases upon the increase of the temperatures. When a
critical temperature $T_{\chi}^c \simeq 201  \mathrm{MeV}$ is reached, $\Delta \varepsilon$ vanishes, which indicates that the chiral symmetry is restored. Moreover, according
to the above criterion for the existence of the stable soliton solution, for zero chemical potential, we can find the stable soliton solutions at various temperature from zero
temperature until to the critical temperature for the chiral phase transition $T_{\chi}^c$. The result is believed to be held for all crossover transition region in
the QCD phase diagram.

For $\mu=380 \mathrm{MeV}$, one clearly observes the characteristic pattern of a first order phase transition: two minima corresponding to phases of restored and broken
symmetry are separated by a potential barrier and they will become degenerate at $T=T_{\chi}^c$. Chiral symmetry is approximately restored for $T>T_{\chi}^c$, where the
minimum at perturbative vacuum $\sigma\sim 0$ becomes the absolute minimum as shown in the right panel in Fig.\ref{Fig02}. The bag constant $B$ is now negative, then it is
physically prohibited to support the existence of the stable soliton solution, so that there is no soliton solution anymore. Therefore, we can only obtain the stable soliton
solution for $T\leq T_{\chi}^c$, and this is also applicable for the whole first-order transition region in the QCD phase diagram. Moreover, the barrier between
the two local minima of the effective potential around $T_{\chi}^c$, shown in the right panel in Fig.\ref{Fig02}, will decrease with decreasing $\mu$. At a specific chemical
potential, $\mu^c$, the barrier will finally disappear and the transition will become of second order. The point $C$ ($T_{\chi}^c$, $\mu^c$) of the phase diagram will be
termed as the critical end point (CEP).

\begin{figure}[thbp]
\epsfxsize=9.0 cm \epsfysize=6.5cm
\epsfbox{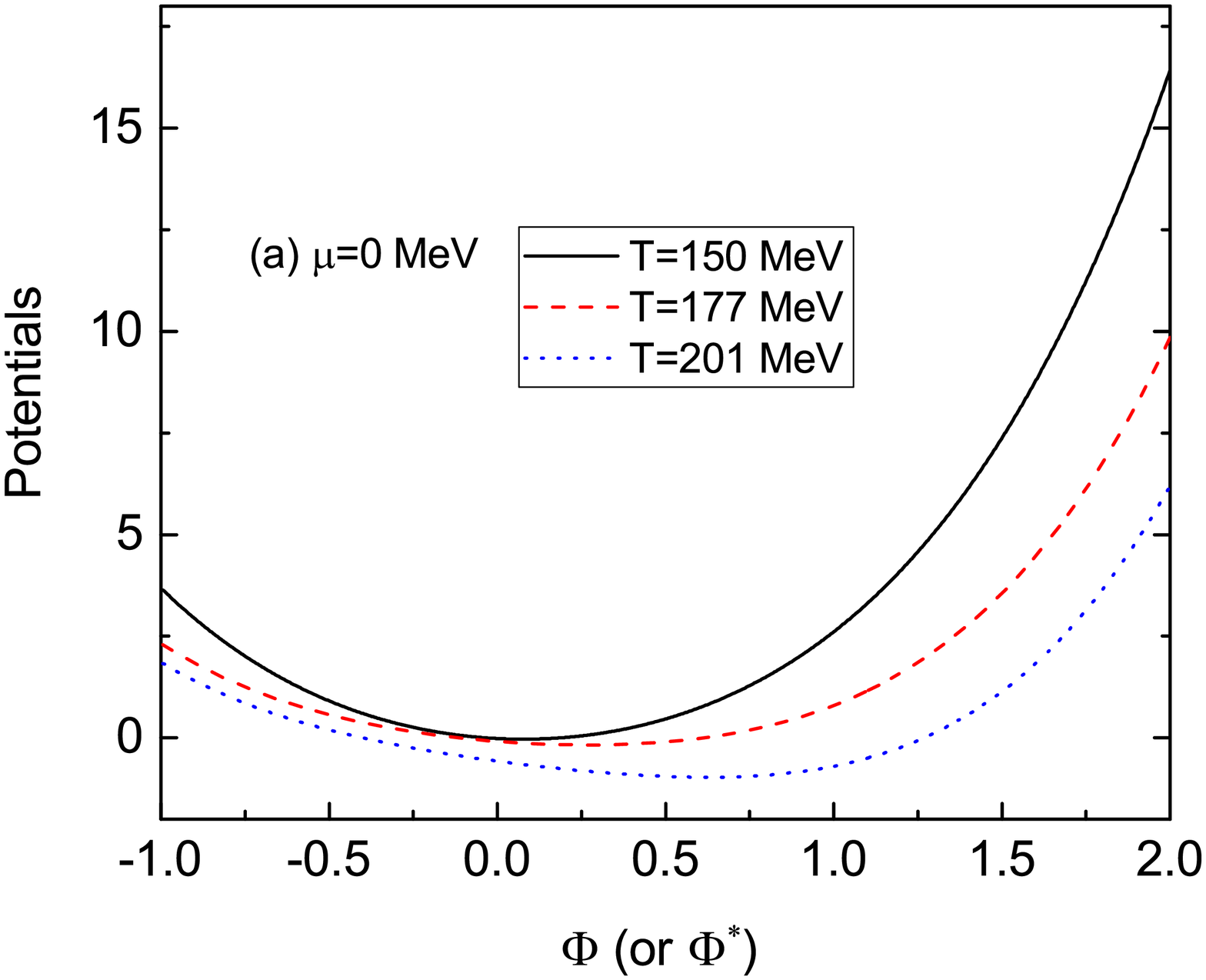}\hspace*{0.1cm} \epsfxsize=9.0 cm
\epsfysize=6.5cm \epsfbox{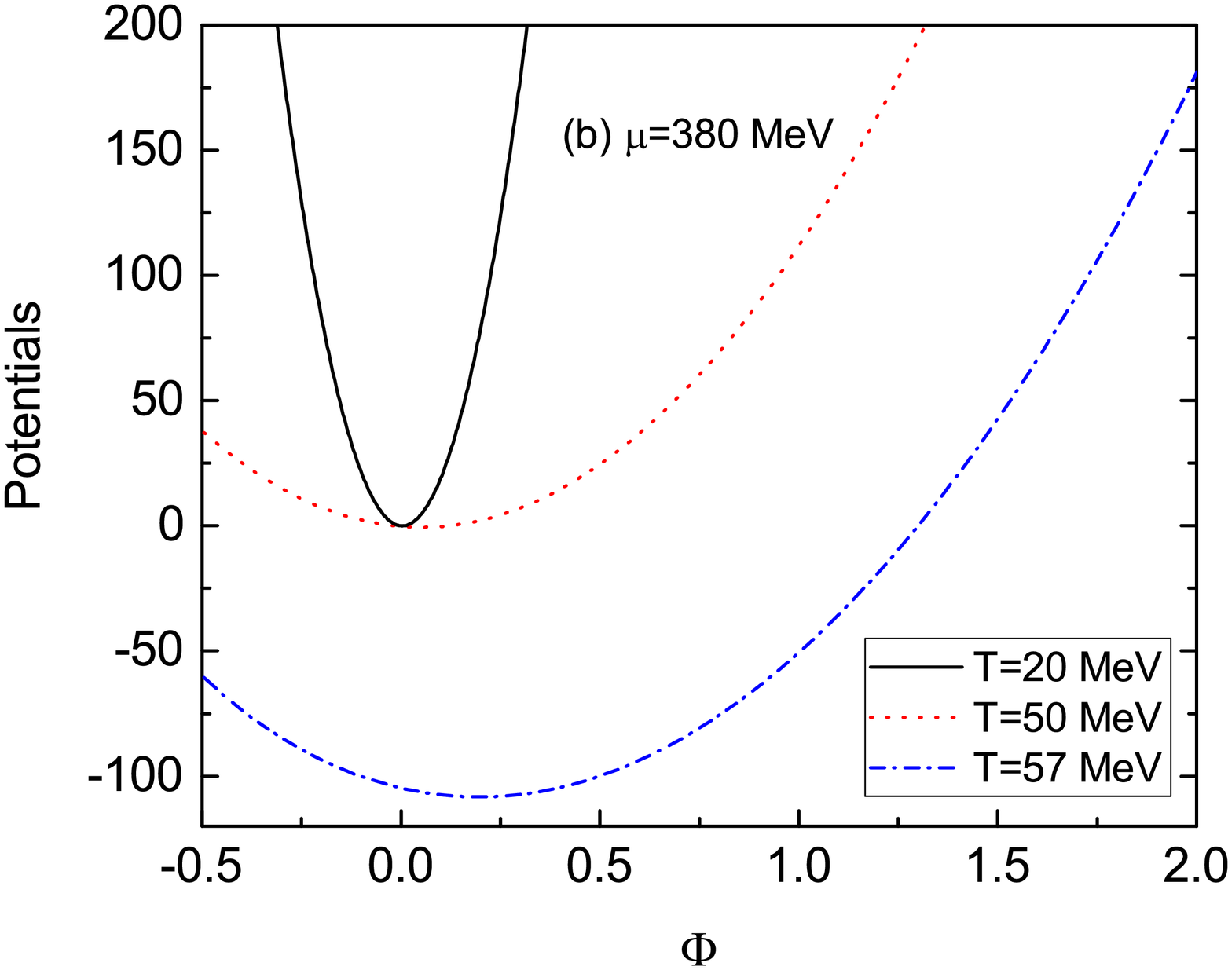}
%\vskip -0.05cm \hskip 0.15 cm \textbf{( a ) } \hskip 6.5 cm \textbf{( b )} \\
 \caption{(Color online) (a) The grand canonical potential $\Omega_{\mathrm{MF}}$ as a function of the Polyakov loop $\Phi$ (or $\Phi^*$) for $\mu=0 \mathrm{MeV}$ by fixing the chiral order
parameters on their expectation values. (b)The grand canonical potential $\Omega_{\mathrm{MF}}$ as a function of the Polyakov loop $\Phi$ for $\mu=380 \mathrm{MeV}$ by fixing
the chiral order parameter on their expectation values. $\Omega_{\mathrm{MF}}$ is scaled by a factor of $T^4$. }
\label{Fig03}
\end{figure}

Let us now investigate how the grand canonical potentials $\Omega_{\mathrm{MF}}$ evolve with the Polyakov loop field $\Phi$ for different chemical potential, by fixing the
chiral order parameters on their expectation values. The scaled grand canonical potential is shown in Fig.\ref{Fig03} as a function of $\Phi$. From Fig.\ref{Fig03} it is
obvious that these effective potentials share similar behaviors for both $\mu=0 \mathrm{MeV}$ and $\mu=380 \mathrm{MeV}$: there is only one minimum for each of the
effective potentials, and these minima correspond to the expectation values $\bar{\Phi}$ of the Polyakov loop field at specific temperature and density. In the crossover
transition region, with the raising of the temperature, the expectation value $\bar{\Phi}$ moves to its higher value smoothly and slowly. But in the first-order transition
region at high density, accompanied by the jump of chiral order parameter $\sigma$, the expectation value $\bar{\Phi}$ develops a disconnection across the gap from relatively
small value to its maximum, which indicates that there exists a degenerate value of the Polyakov loop variable $\Phi$ along the first-order transition line in the QCD phase
diagram for very high chemical potential. This implies that the integration of the quark and meson fields in the grand canonical potential $\Omega_{\mathrm{MF}}$ would only
result in a trivial Polyakov loop effective potential at finite $T$ and $\mu$, and such a naive potential does not tell anything about the critical point at which the
deconfinement transition should definitely happened. Therefore, the jump induced by the chiral order parameter is not to be supported by the effective potential of the
Polyakov loop field itself, then it certainly can not be treated as a signal for the deconfinement phase transition. This is the reason why we argue that there exists no
obvious criterion for defining the critical temperature for the deconfinement phase transition in terms of using the Polyakov loop variables ${\Phi}$, ${\Phi}^*$
or their temperature derivatives, as long as the temperature $T$ is smaller than the critical temperature $T_0$ for deconfinement in the pure gauge sector.

Nevertheless, the advantage is that we do not have a bag-like soliton solutions for the Polyakov loop variables ${\Phi}$, ${\Phi}^*$, since there is only one minimum in the
effective Polyakov loop potential. On the other hand, the Polyakov loop variables ${\Phi}$, ${\Phi}^*$ will always develop their expectation values $\bar{\Phi}$ and
$\bar{\Phi}^*$ in whole space, such that these fields should be regarded as homogeneous background thermal fields on top of which the chiral soliton is to be added on.

\section{ Nontopological soliton solution in the model}
In vacuum, the Polyakov loop variables ${\Phi}$, ${\Phi}^*$ are set to zero and the thermodynamic grand potential $\Omega_{\mathrm{MF}}$ reduces to the purely mesonic
potential $\Omega_{\mathrm{M}}$. Following reference \cite{Mao:2013qu}, in the mean field approximation, the $\sigma$ and $\pi$ are taken as time-independent, classical
$c$-number fields, which only differ from their vacuum values in the neighborhood of the quark sources. The state of the quarks $\{\phi_n(\mathbf{r}) \}$ with energy
$\{\epsilon_n\}$ and the $\sigma(\mathbf{r})$, $\pi(\mathbf{r})$ meson fields satisfy the coupled set of the Euler-Lagrange equations of motion
\begin{eqnarray}
-i \vec{\alpha}\cdot \vec{\nabla}\phi_n(\mathbf{r})-g \beta \left[\sigma(\mathbf{r})+i \gamma_5 \vec{\tau}\cdot \vec{\pi}(\mathbf{r}) \right] \phi_n(\mathbf{r})=\epsilon_n
\phi_n(\mathbf{r}), \label{eom1}\\
-\nabla^2 \sigma (\mathbf{r})+\frac{\partial \Omega_{\mathrm{M}}(\sigma_v)}{\partial \sigma}=-g \sum_{n_{occ}} \bar{\phi}_n(\mathbf{r}) \phi_n(\mathbf{r}) \label{eom2}\\
-\nabla^2 \vec{\pi} (\mathbf{r})+\frac{\partial \Omega_{\mathrm{M}}(\sigma_v)}{\partial \vec{\pi}}=-g \sum_{n_{occ}} \bar{\phi}_n(\mathbf{r})i \gamma_5 \vec{\tau}
\phi_n(\mathbf{r})
\label{eom3}
\end{eqnarray}
with
\begin{eqnarray}
\int \phi^{\dag}_n(\mathbf{r}) \phi_n(\mathbf{r}) d^3 r=1
\end{eqnarray}
where $\vec{\alpha}$ and $\beta$ are the conventional Dirac matrices.

The ground state of the chiral soliton is the state with $N$ quarks in the same lowest Dirac state $\phi_0$, with energy $\epsilon$. In the following, our discussions are
constrained in the case of $N=3$ for baryons. In order to obtain solutions of minimum energy, we adopt the ``hedgehog" ansatz with the meson fields are spherically symmetric
and valence quarks are in the lowest s-wave level
\begin{eqnarray}
\sigma = \sigma(r), \vec{\pi} = \hat{\mathbf{r}}\pi(r),\\
\phi_0 = \left(\begin{array}{c} u(r) \\
i \vec{\sigma} \cdot\mathbf{\hat{ r}}v(r)
\end{array}\right)\chi, \label{sltconf}
\end{eqnarray}
where $\chi$ is a state in which the spin and isospin of the quark couple to zero:
\begin{eqnarray}
(\vec{\sigma}+\vec{\tau})\chi=0.
\end{eqnarray}

\begin{figure}
\includegraphics[scale=0.36]{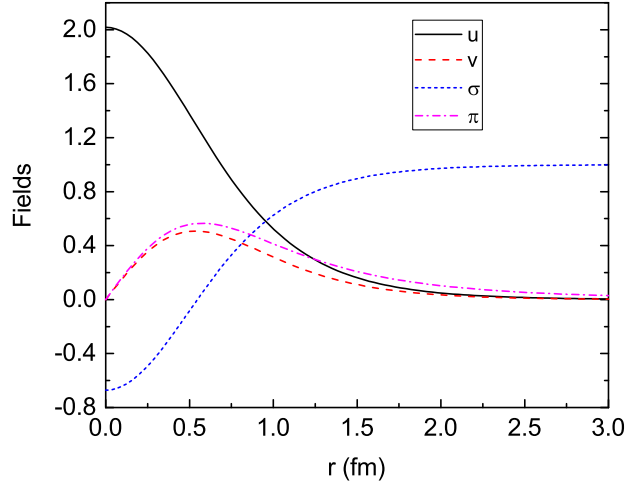}
\caption{\label{Fig12}(Color online) The quark fields in relative unit and the $\sigma$, $\pi$ fields scaled with $f_{\pi}$ as function of the radius $r$ in vacuum.}
\end{figure}

Now the system is spherical symmetric and the Euler-Lagrange equations of motion (\ref{eom1})-(\ref{eom3}) transform in radial coordinates to
\begin{eqnarray}
\frac{du(r)}{dr} =  -\left(\epsilon+g \sigma(r)\right)v(r)-g \pi(r)u(r), \label{equation1}\\
\frac{dv(r)}{dr} =  -\left(\frac{2}{r}-g\pi(r)\right)v(r)+\left(\epsilon-g \sigma(r)\right)u(r), \label{equation2} \\
\frac{d^2 \sigma(r)}{dr^2}+\frac{2}{r}\frac{d\sigma(r)}{dr}-\frac{\partial \Omega_{\mathrm{M}}}{\partial \sigma}= Ng\left(u^2(r)-v^2(r)\right),\label{equation3}\\
\frac{d^2 \pi(r)}{dr^2}+\frac{2}{r}\frac{d\pi(r)}{dr}-\frac{2 \pi(r)}{r^2}-\frac{\partial \Omega_{\mathrm{M}}}{\partial \pi} = -2Ng u(r) v(r),
\label{equation4}
\end{eqnarray}
and the quark functions should satisfy the normalization condition
\begin{eqnarray}\label{norm}
4\pi \int r^2 \left(u^2(r)+v^2(r)\right)dr=1.
\end{eqnarray}
These equations are subject to the boundary conditions which follow
from the requirement of finite energy:
\begin{eqnarray}
v(0)=0,     \frac{d\sigma(0)}{dr}=0,\pi(0)=0 \label{boundary1}\\
u(\infty)=0,\sigma(\infty)={\it f}_{\pi},\pi(\infty)=0.\label{boundary2}
\end{eqnarray}
The asymptotic vacuum value of the soliton field has to be determined by an additional condition, i.e. that the physical vacuum is recovered at infinity. In this ``physical"
vacuum the quarks are free Dirac particles with the constituent mass $g \sigma_v$, and the chiral symmetry is spontaneously broken. By solving the coupled differential
equations (\ref{equation1}),(\ref{equation2}),(\ref{equation3}) and (\ref{equation4}) with the normalization and the appropriate boundary conditions, in Fig.\ref{Fig12} we
plot the $\sigma$, $\pi$ and quark fields profiles in arbitrary unit as functions of $r$ for zero temperature and chemical potential.

If we put $N$ quarks into the lowest state with energy $\epsilon$, the total energy of the hedgehog baryon is given by
\begin{eqnarray}\label{energy}
E=N \epsilon+4\pi\int r^2 \left[ \frac{1}{2}
\left(\frac{d\sigma}{dr}\right)^2+\frac{1}{2}\left(\frac{d\pi}{dr}\right)^2+\frac{\pi^2}{r^2}+\Omega_{\mathrm{M}}(\sigma_v) \right]dr,
\end{eqnarray}
which is normally identified as the mass of the nucleon $M_N$ below.

As a next step we consider a $B = 1$ localized bound state (soliton) in a thermal medium. Customarily, the thermal medium can be treated as a quark medium or a nuclear medium
due to the interaction of the three valence quarks with the quark Dirac sea and the nucleon Fermi sea via the meson fields\cite{Berger:1996hc}.

For the quark medium, the thermal medium is filled with quarks of a constituent mass $M_q$, and the soliton energy is given by the sum of the energy of the valence quarks, the
meson fields and their interactions as shown in Eq.(\ref{energy}). Then a new set of coupled equations of motion for the meson fields could be derived by simply replacing the
relevant mesonic potential $\Omega_{\mathrm{M}}$ with the thermodynamic grand potential $\Omega_{\mathrm{MF}}$. Accordingly, a set of coupled equations for mesons can be
described as
\begin{eqnarray}
\frac{d^2 \sigma(r)}{dr^2}+\frac{2}{r}\frac{d\sigma(r)}{dr}-\frac{\partial\Omega_{\mathrm{MF}}}{\partial\sigma} = Ng\left(u^2(r)-v^2(r)\right),\label{equation5}
\label{equation5}\\
\frac{d^2 \pi(r)}{dr^2}+\frac{2}{r}\frac{d\pi(r)}{dr}-\frac{2 \pi(r)}{r^2}-\frac{\partial\Omega_{\mathrm{MF}}}{\partial\pi} = -2Ng u(r) v(r).
\label{equation6}
\end{eqnarray}
For satisfying the requirement of finite energy of the soliton, one of the boundary conditions in Eq.(\ref{boundary2}) should modified accordingly as: $r\rightarrow \infty$,
$\sigma(r)$ approaches to the expectation value $\bar{\sigma}_v$, where thermodynamic grand potential $\Omega_{\mathrm{MF}}$ has an absolute minimum.

As long as the unbound constituent quarks, treated as the homogeneous background thermal fields with $T$ and $\mu$, are allowed to penetrate into the soliton, they
will bring an additional contribution to the total baryon density. Thus to ensure that the solitonic baryon number is equal to one, the normalization condition equation
(\ref{norm}) should be modified as
\begin{eqnarray}\label{norm2}
4\pi \int r^2 \left(u^2(r)+v^2(r)\right)dr=1-B_m,
\end{eqnarray}
with
\begin{eqnarray}
B_m=4\pi \int_V \rho^m_B r^2 dr.
\end{eqnarray}
Here, $\rho^m_B= -\frac{1}{3}\frac{\partial \Omega_{\mathrm{MF}}}{\partial \mu}$ and $V$ being the volume of the soliton.

In contrast, when a soliton is embedded in the medium of nucleons we have to consider a Fermi sea of nucleons instead of quarks, due to confinement. This is because the
Dirac sea consists of quarks and therefore only determines the vacuum sector. Thus in this case the mesons are directly coupled to the nucleons of the Fermi sea. Accordingly,
the terms representing the thermal medium effects in Eqs.(\ref{equation5}\ref{equation6}) should be modified as:
\begin{eqnarray}
\frac{\partial\Omega_{\mathrm{MF}}}{\partial\sigma} &\rightarrow & g_N \langle \bar{\psi}_N\psi_N\rangle, \\
\frac{\partial\Omega_{\mathrm{MF}}}{\partial\pi} &\rightarrow & g_N \langle \bar{\psi}_N i \gamma_5 \vec{\tau}\psi_N\rangle.
\end{eqnarray}
Here, the bracket $\langle \rangle$ denotes the expectation value of the operator between the nuclear ground state, $\psi_N$ is the nucleon field and $g_N$ is a coupling
constant which relates the nucleon mass to the non-zero expectation value of the scalar meson field. Unluckily, both the coupling constant $g_N$ and the scalar and
pseudoscalar densities of nucleons and antinucleons in the above equations cannot be obtained from the present model self-consistently. In fact, they have to be considered as
input parameters which could be taken from the RMF approximation or the BHF theory \cite{Ring:1980}\cite{Serot:1986}. Therefore, in order to avoid inconsistencies, it is
customary to treat a hot and dense thermal medium as a uniform constituent quark and gluon medium (quark medium) with solitons embedded, as in
\cite{Mao:2013qu}\cite{Berger:1996hc}\cite{Schleif:1997pi}.

As long as there is no bag-like soliton solution for the Polyakov loop variables ${\Phi}$, ${\Phi}^*$ in whole space, the Polyakov loop variables ${\Phi}$, ${\Phi}^*$ take
constantly their expectation values $\bar{\Phi}$ and $\bar{\Phi}^*$. Hence these variables denote the contributions only to the thermodynamic grand potential
$\Omega_{\mathrm{MF}}$ rather than to the equations of motion for the nontopological soliton solutions. Consequently, the properties of a soliton placed in a thermal medium
can be investigated by solving the four coupled Euler-Lagrange equations that arise from the thermodynamic grand potential $\Omega_{\mathrm{MF}}$ in Eq.(\ref{omegamf}). This
system of equations does not possess analytic solutions, but is readily solved numerically. Various numerical packages are available for the solution of such equations. One
that has been widely used in this field, is COLSYS\cite{Ascher:1981}.

\section{Nucleon static properties at finite temperature and density}
\begin{figure}[thbp]
\epsfxsize=9.0 cm \epsfysize=6.5cm
\epsfbox{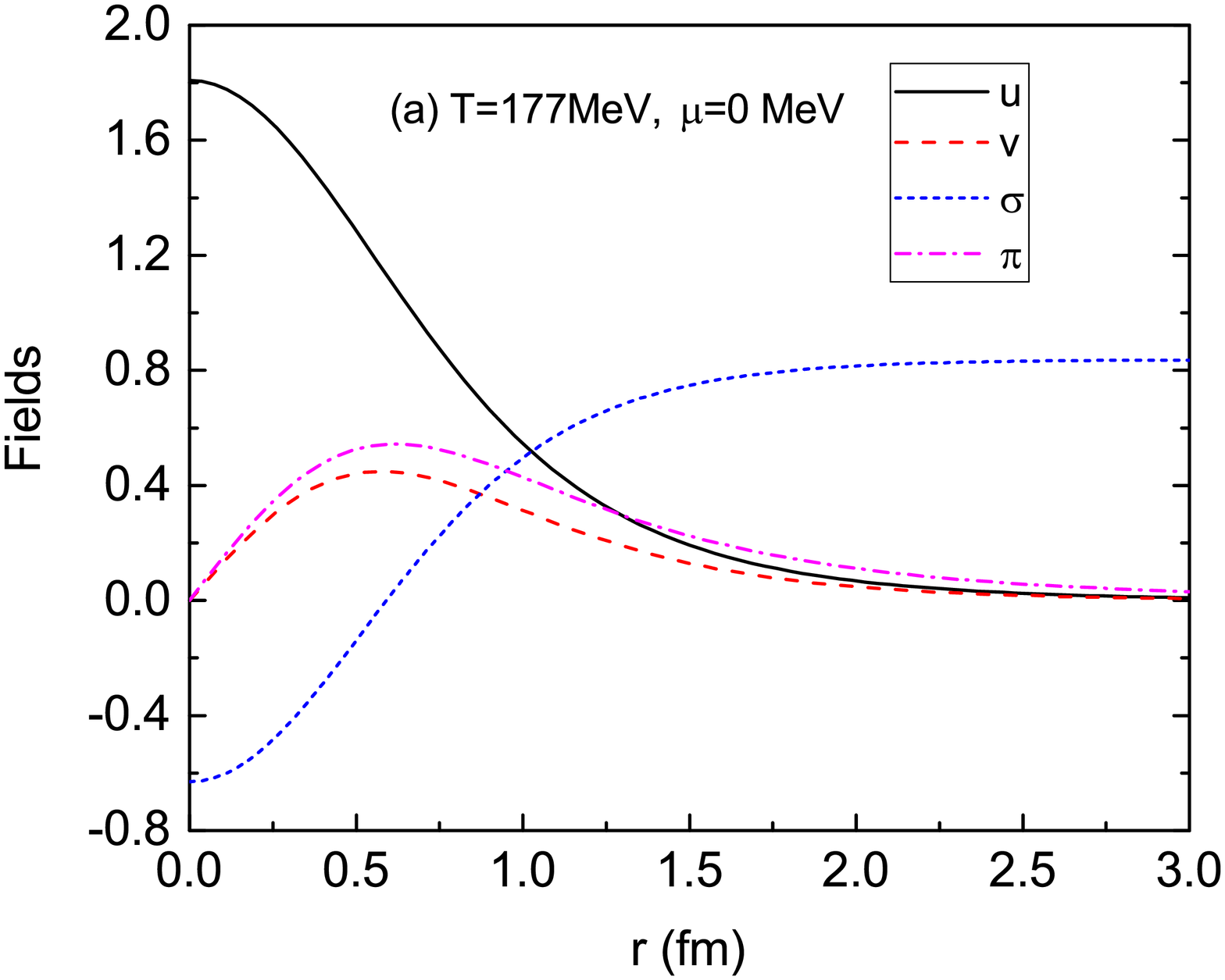}\hspace*{0.1cm} \epsfxsize=9.0 cm
\epsfysize=6.5cm \epsfbox{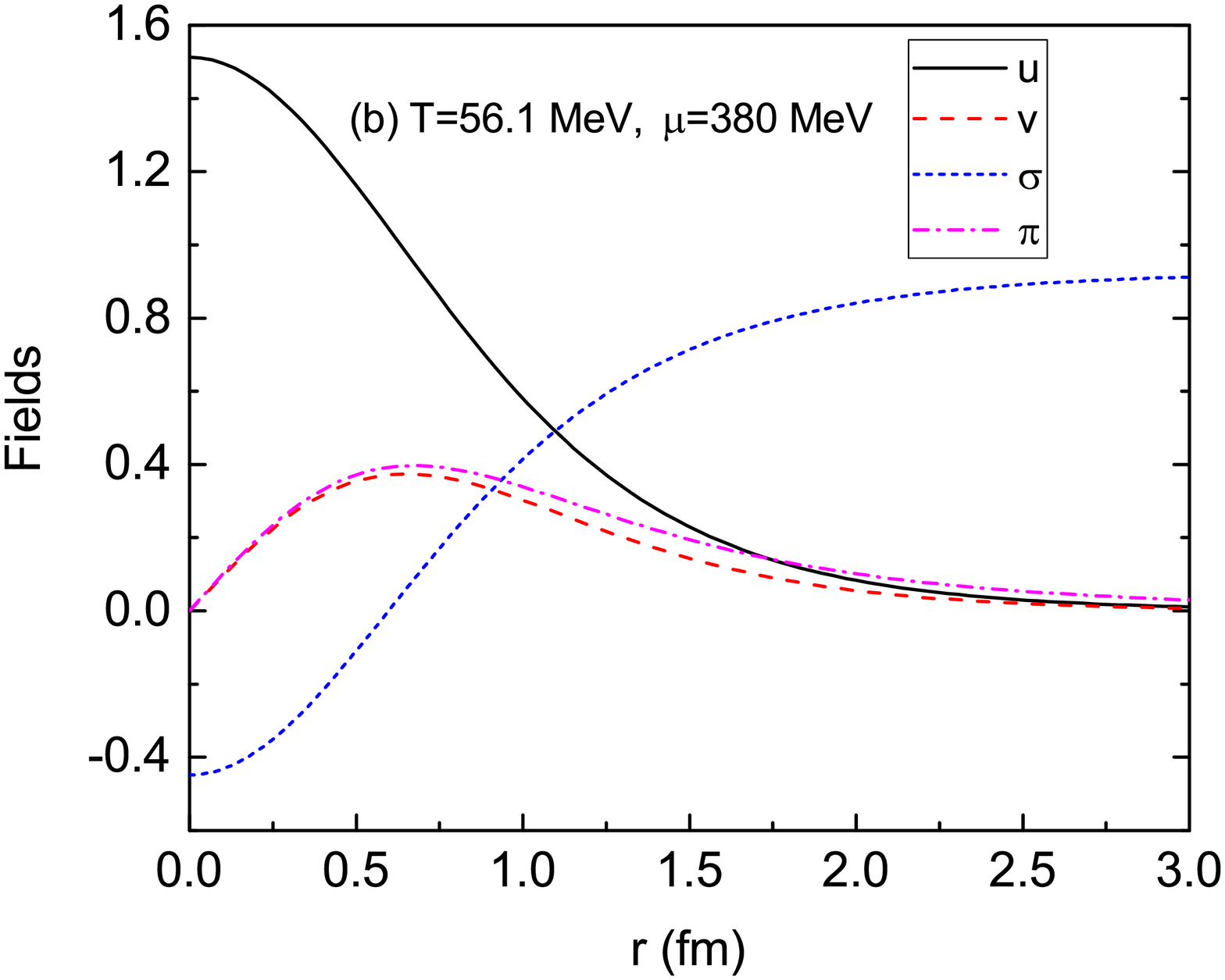}
%\vskip -0.05cm \hskip 0.15 cm \textbf{( a ) } \hskip 6.5 cm \textbf{( b )} \\
 \caption{ (Color online) (a) The quark fields in relative units and the $\sigma$, $\pi$ fields scaled with $f_{\pi}$ as function of the radius $r$ for $T=177 \mathrm{MeV}$ as $\mu=0
\mathrm{MeV}$. (b) The quark fields in relative units and the $\sigma$, $\pi$ fields scaled with $f_{\pi}$ as function of the radius $r$ for $T=56.1 \mathrm{MeV}$ while
$\mu=380 \mathrm{MeV}$.}
\label{Fig04}
\end{figure}

We firstly study soliton solutions at finite temperature and density by solving the coupled differential equations (\ref{equation1}),(\ref{equation2}),(\ref{equation5}) and
(\ref{equation6}) with the normalization condition and the appropriate boundary conditions. In Fig.\ref{Fig04}, we plot the $u(r)$, $v(r)$, $u(r)$ and $v(r)$ fields at zero
and finite chemical potential ($\mu=380$ MeV) for different temperatures. These two chemical potentials correspond to the typical crossover and first-order phase transitions
in the QCD phase diagram, respectively. For both cases, it is shown that all the fields are moving towards the trivial values while the temperature increasing. When $T$ is
lager than some critical temperature $T_{\chi}^c$, there only exist trivial solutions for the coupled equations of motion and solitons are melted away. These trivial solutions
indicate the restoration of the chiral symmetry in full space. Moreover, the lack of solitonic solutions are usually considered as a signal for the delocalization of the
baryonic phase.

\begin{figure}
\includegraphics[scale=0.36]{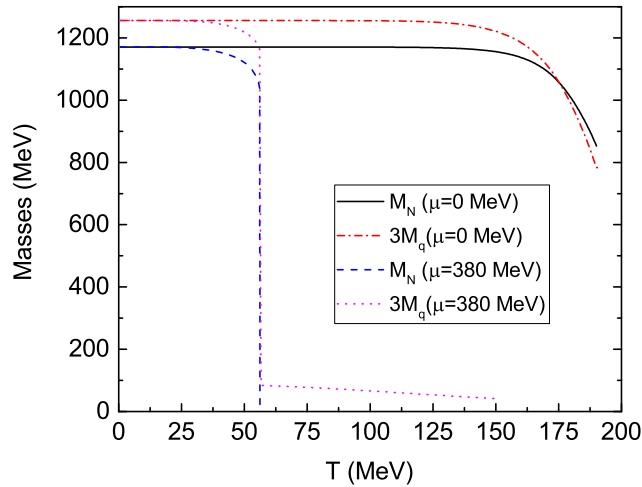}
\caption{\label{Fig05}(Color online) The total energy of system $M_N$ and the energy of $3$ free constituent quark $3M_q$ as function of the temperature $T$. Here one set is for $\mu=0
\mathrm{MeV}$, another set is for $\mu=380 \mathrm{MeV}$.}
\end{figure}

Based on the above analysis, for both the crossover and first-order transitions, the effective potential supports the existence of the stable soliton solution for the meson
fields, as long as $T$ is lower than $T_{\chi}^c$. This implies that the baryonic phase can be indeed found in the chiral symmetry-breaking phase. However, the
stability of such baryonic phase should be checked carefully by comparing the total energy of the system in thermal medium with the energy of three free constituent quarks in
the system. By subtracting the homogeneous medium contribution\cite{Mao:2013qu}\cite{Berger:1996hc}, the total energy of system $M_N$ is plotted as a function of the
temperature for $\mu=0 \mathrm{MeV}$ and $\mu=380 \mathrm{MeV}$ in Fig.\ref{Fig05}. Here, one finds that for a smooth crossover of the symmetry breaking patten at low density
region, both $M_N$ and $3M_q$ fall smoothly from the corresponding vacuum value as $T$ goes to high temperature. When $T$ is close to some high critical temperature
$T_{\chi}^c\ simeq 201 \mathrm{MeV}$ for $\mu=0 \mathrm{MeV}$, both $M_N$ and $3M_q$ experience a steep descent region. But as shown in Fig.\ref{Fig05}, in the high
temperature regime $3M_q$ drops more quickly than that of $M_N$, as $T>177 MeV$, $3M_q<M_N$. This implies that even though the stable soliton solution still exists
in the temperature region $T\in [177,201] \mathrm{MeV}$, it is energetically unfavorable. The baryonic phase will definitely melt away into three free constituent quarks. In
this way, we can identify this specific temperature $T_d^c\simeq 177\mathrm{MeV}$ as a critical temperature of the deconfinement phase transition.

Based on the right panel in Fig.\ref{Fig05}, one can see that in the large density region the total energy $M_N$ decreases monotonically with increasing the temperature $T$
from zero to higher values. As the temperature approaches the critical temperature $T_{\chi}^c$, $M_N$ starts to deviate from the ones in vacuum significantly. When
$T>T_{\chi}^c$, $M_N$ jumps to zero quickly, which indicates the delocalization phase transition from nucleon matter to quark matter due to the fact that the effective
potential does not support the existence of the stable soliton solution. The energy of three free constituent quark $3M_q$ (or $\sigma_v$) shows the similar behavior as
$M_N$. By comparing the two energies in Fig.\ref{Fig05}, we can show that for $T<T_{\chi}^c$ the nucleon bound sate is stable and $3E_q$ is larger than $M_N$, but the
difference decreases with the increase of temperature, and the two energies begin to cross over at the critical temperature $T_{\chi}^c$. Therefore, the critical temperature
for the deconfinement phase transition is coincident with that of the chiral phase transition for the first-order phase transition.

\begin{figure}
\includegraphics[scale=0.36]{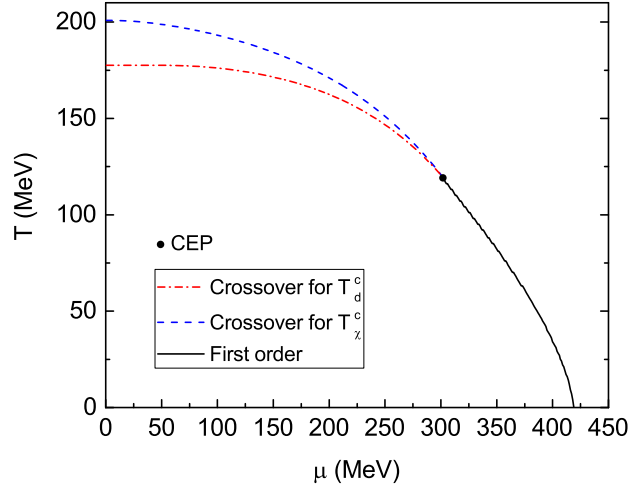}
\caption{\label{Fig06}(Color online) Two-flavor phase diagram in the $T-\mu$ plane in the Polyakov quark meson model based on the nontopological soliton picture. The dash-dotted curve is
the critical line for $T_d^c$ which characterizes the confinement phase transition, and the dashed lines are the critical line for conventional chiral phase transition in the
region of crossover. The solid lines indicates the first-order phase transitions, and the solid circle indicates the critical end points for chiral phase transitions of $u$
and $d$ quarks.}
\end{figure}

We infer the occurrence of the chiral phase transitions of $u$ and $d$ quarks and the deconfinement phase transition at finite temperature and finite density, and show the
$T-\mu$ phase structure of the Polyakov quark meson model in Fig.\ref{Fig06} based on the nontopological soliton picture. For two light flavors, there is a crossover in the
low density region and a first-order phase transition in the high-density region, and in the middle position there exists a critical end point (CEP). From
figure \ref{Fig06}, the critical temperature for the deconfinement phase transition $T_d^c$ is lower than that of the chiral phase transition, and both two critical
temperatures decrease smoothly as $\mu$ goes to high value. With the increasing of $\mu$, the difference between $T_{\chi}^c$ and $T_d^c$ becomes smaller and smaller, while at
some critical chemical potential $\mu^c$ it becomes zero, which identifies the critical end point (CEP) for a second order phase transition. The corresponding values are
$(T^c, \mu^c)\simeq (119, 302)\mathrm{MeV}$.

Here are several remarks on the phase diagram presented in Fig.\ref{Fig06}. From the above discussions on the effective potential and nontopological soliton concerning the
deconfinement and chiral phase transition, we conclude that the effective potential always support the existence of the stable soliton solution in the system if $T\leq
T_{\chi}^c$, but not for $T> T_{\chi}^c$. It is required that the critical temperature defined as the deconfinement phase transition in a nontopological soliton model usually
less than the critical temperature for the chiral phase transition, as $T^c_d\leq T_{\chi}^c$. In this study, in the first-order region, we take the ``$=$''. In contrast, in
the crossover region we should have the ``$<$''. This conclusion is in qualitative agreement with the result shown in Fig.6 in Ref.\cite{Gupta:2011ez} at relatively
low and middle densities. But for high density, they produce a very strange behavior for the deconfinement crossover phase transition for the Polyakov loop variables
${\Phi}$, ${\Phi}^*$. This is different from the Friedberg-Lee model\cite{Reinhardt:1985nq, Gao:1992zd, Li:1987wb} and its descendant models\cite{Mao:2013qu}\cite{Mao:2006zp},
which only predict a first-order phase transition in the phase diagram. The chiral soliton model combined with the Polyakov loop has an obvious advantage in the description
of QCD phase diagram, since it can allow the prediction of the crossover transition at low and middle chemical potential. Nonetheless, it shows the first-order phase
transition for high chemical potential. The result is in agreement with other predictions demonstrated in effective models and lattice QCD
data\cite{Rischke:2003mt,Yagi:2005yb,Fukushima:2010bq}.

\begin{figure}
\includegraphics[scale=0.36]{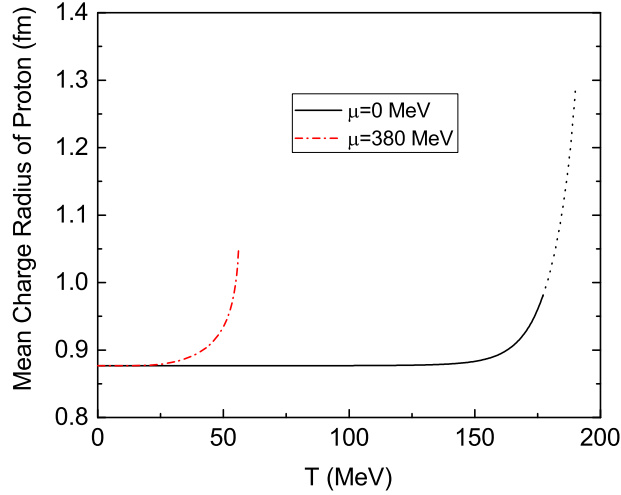}
\caption{\label{Fig07}(Color online) The proton charge `r.m.s' radius of a stable chiral soliton as a function of temperature $T$ at $\mu=0$ MeV and $\mu=380$ MeV. The solid curve is for $\mu=0$ MeV while the dash-dotted curve for $\mu=380$ MeV. The dotted curve is for the unstable baryonic phase existing in the crossover phase transition when $T_{\chi}^c\geq
T>T_d^c$.}
\end{figure}

At the end of this section, the proton charge r.m.s radii $R$ of a stable chiral soliton as a function of temperature for $\mu=0$ MeV and $\mu=380$ MeV are illustrated in
Fig.\ref{Fig07}, it gives a signal of a swelling of the nucleon when temperature and density increase. In both cases, $R$ increases slightly at low temperatures
while the latter is increased. As $T$ approaches $T_d^c$, $R$ sharply grows and disappears. Another interesting result displayed in Fig.\ref{Fig07} is that the maximal
radius $R$ at various densities are almost same when $T$ near $T_d^c$, which hints that solitons start to overlap each other with the similar expansion rate at $T_d^c$ for
different densities.

\section{QCD thermodynamics at zero chemical potential}
In order to investigate the influence of the Polyakov loop on the equilibrium thermodynamics of the system, we calculate the pressure of the system $P$ during the QCD phase
transition form hadron phase to quark phase according to two different models as follows.

Firstly, we adopt the mean-field approximation as usual by replacing $\sigma$, $\vec{\pi}$ and the Polyakov loop variables ${\Phi}$, ${\Phi}^*$ by their expectations values.
In other words, we neglect both quantum and thermal fluctuations of the meson fields and the Polyakov loop variables but retain the quarks and antiquarks as only quantum
fields in the entire phase diagram. This is of course not a realistic scheme, especially at low $T$ and $\mu$, since due to the confining forces quark and antiquarks will
recombine into mesons, baryons and antibaryons. Hence, the character of the chiral phase transition described by the mean-field approximation could be drastically changed in
hadronization process from quark phase to hadron phase. But, if we discard these affects, all thermodynamic quantities can be obtained from the grand canonical potential in a
spatially uniform system $\Omega_{\mathrm{MF}}$ in Eq.(\ref{omegamf}), which is determined as the logarithm of the partition function. The negative of grand potential
which is normalized to vanish at $T=\mu=0$ gives the thermodynamic pressure in PQM model
\begin{eqnarray}
P_{\mathrm{PQM}}=-\Omega_{\mathrm{MF}}(T,\mu).
\end{eqnarray}
The pressure obtained in the above equation can be directly compared with lattice data.

However, the hadron and quark phases can be distinguished by empirical facts and phenomena at low and high energies. At low temperature and low baryon density, the hadronic
phase exhibits a dynamical breaking of chiral symmetry and the confinement, and the baryon and meson act as the active degrees of freedom here. On the contrary, at very high
temperature or baryon density, quarks and gluons will be set free to play the dominant roles in quark gluon plasma. Such a scenario can be realized in the nontopological
soliton (NS) model vividly as follows: in the hadron phase, the state of the free quarks is not the ground state of the strongly interacting matter, and as a result three
valence quarks will form the bound state of the nucleon. Therefore, the hadron phase only possesses baryons and mesons. On the other side, when $T>T^c_d$, the solitons are
going to dissolve, and the hadronic phase will eventually evolve to quark phase with free quarks.

For simplicity, within the NS model, we assume an ideal case of the system by taking the hadronic phase as a noninteracting hadron gas composed of nucleons and
$\pi$, $\sigma$ mesons with the effective masses $M_N$, $M_{\pi}$ and $M_{\sigma}$ in thermal medium. The Polyakov-loop variables are treated as the background thermal fields
and accordingly the Polyakov-loop potential has been subtracted already. It is then straightforward to write down the normalized pressure of the system in terms of nucleons
and mesons for the hadronic phase\cite{Yagi:2005yb}\cite{Kapusta:2006pm}
\begin{eqnarray}
P^{\mathrm{H}}_{\mathrm{NS}} &=& \nu_N T \int \frac{d^3\vec{p}}{(2
\pi)^3} \left\{ \mathrm{ln} \left[ 1+e^{-(E_N-\mu_B)/T} \right] +\mathrm{ln} \left[
1+e^{-(E_N+\mu_B)/T}\right]\right\} \nonumber \\&&
-\nu_{\pi} T \int \frac{d^3\vec{p}}{(2
\pi)^3} \left\{ \mathrm{ln} \left[ 1-e^{-E_{\pi}/T}\right]\right\}-\nu_{\sigma} T \int \frac{d^3\vec{p}}{(2
\pi)^3} \left\{ \mathrm{ln} \left[ 1-e^{-E_{\sigma}/T}\right]\right\}-B^*(M_N).
\label{pressure}
\end{eqnarray}
Where $\nu_N=4$ for nucleon, $\nu_{\pi}=3$ for pion and $\nu_{\sigma}=1$ for sigma meson. The last term $B^*(M_N)$ is introduced in order to recover the thermodynamical
consistency of the system, since the nucleons are treated as the chiral solitons with a temperature-dependent masses\cite{Gorenstein:1995vm}. The explicit expression of this
term can be evaluated by the additional constraint $(\partial P_{HP}/\partial M_N)_T=0$, which gives
\begin{eqnarray}
B^*\left(M_N(T) \right) = B^*\left(M_N(0) \right) -\nu_N \int_0^T dT' \frac{d M_N(T')}{dT'} M_N(T') \int \frac{d^3\vec{p}}{(2
\pi)^3} \frac{1}{E_N'}\left[ \frac{1}{e^{(E_N'-\mu_B)/T'}+1} +\frac{1}{e^{(E_N'+\mu_B)/T'}+1} \right],
\end{eqnarray}
with $E_N'=\sqrt{\vec{p}^2+{M_N(T')}^2}$. The energies in Eq.(\ref{pressure}) $E_N=\sqrt{\vec{p}^2+{M_N(T)}^2}$, $E_{\pi}=\sqrt{\vec{p}^2+{M_{\pi}(T)}^2}$ and
$E_{\sigma}=\sqrt{\vec{p}^2+{M_{\sigma}(T)}^2}$ are corresponding to nucleon, pion and sigma mesons, respectively. $M_N$ is obtained as the energy of soliton, whereas, the
$\sigma$ and $\pi$ masses are determined by the curvature of $\Omega_{\mathrm{MF}}$ in Eq.(\ref{omegamf}) at the global minimum:
\begin{eqnarray}
M^2_{\sigma}=\frac{\partial^2 \Omega_{\mathrm{MF}}}{\partial \sigma^2},  M^2_{\pi}=\frac{\partial^2 \Omega_{\mathrm{MF}}}{\partial \pi^2}.
\end{eqnarray}

Since it is unfavorable for solitonic nucleons to survive at high energy when the temperature is across the deconfinement critical temperature $T_d^c \sim 177 \mathrm{MeV}$,
the baryonic bound state formed by three constituent quarks will definitely dissolve and the system should now be regarded as a quark phase including the free quarks,
mesons and the gluons mimicked by the Polyakov loop. Consequently, the pressure of the NS model in terms of free quarks and mesons incorporating with the Polyakov-loop
potential in quark phase results in
\begin{eqnarray}
P^{\mathrm{Q}}_{\mathrm{NS}} &=& \nu_q T \int \frac{d^3\vec{p}}{(2
\pi)^3} \left\{ \mathrm{ln} \left[ 1+e^{-(E_q-\mu)/T} \right] +\mathrm{ln} \left[
1+e^{-(E_q+\mu)/T}\right]\right\} \nonumber \\&&
-\nu_{\pi} T \int \frac{d^3\vec{p}}{(2
\pi)^3} \left\{ \mathrm{ln} \left[ 1-e^{-E_{\pi}/T}\right]\right\}-\nu_{\sigma} T \int \frac{d^3\vec{p}}{(2
\pi)^3} \left\{ \mathrm{ln} \left[ 1-e^{-E_{\sigma}/T}\right]\right\}-\mathbf{\mathcal{U}}(\Phi,\Phi^*,T).
\label{pressure2}
\end{eqnarray}
Here, $\nu_q=2N_cN_f=12$ is the number of internal degrees of freedom of the quarks and $E_q=\sqrt{\vec{p}^2+M_q^2}$ is the valence quark and antiquark energy for $u$ and $d$
quarks, and the constituent quark (antiquark) mass $M_q$ is given by $M_q=g \vec{\sigma}_v$.

\begin{figure}
\includegraphics[scale=0.36]{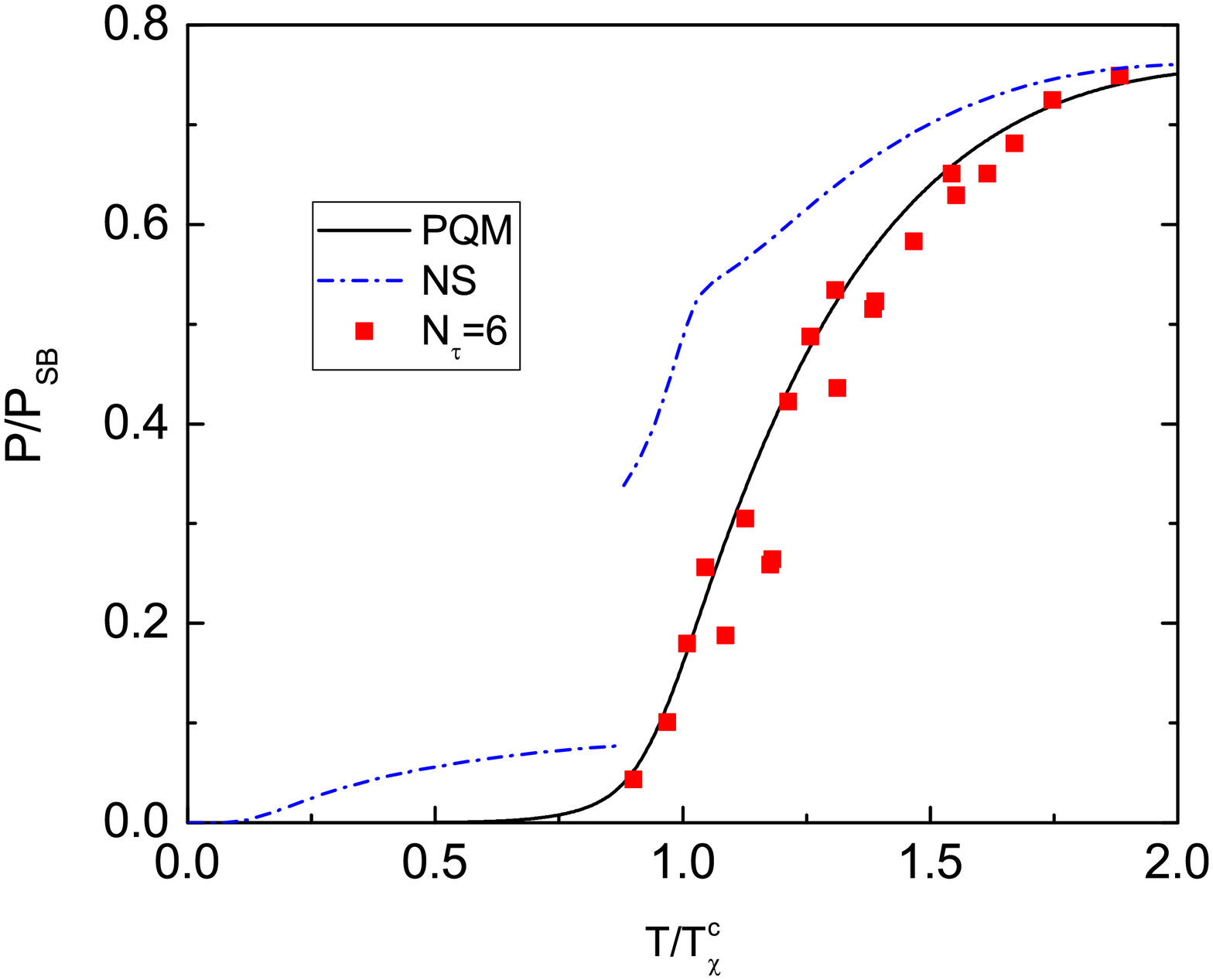}
\caption{\label{Fig08}(Color online) The normalized pressures variation with respect to temperature for PQM model and NS model at $\mu=0$ with $T_{\chi}^c \sim 201 \mathrm{MeV}$. The solid
line corresponds to the pressure in PQM model, while the dash-dotted line is for the pressure in nontopological soliton model.  All calculations are compared to lattice data
($N_{\tau}=6$) from Ref.\cite{AliKhan:2001ek}.}
\end{figure}

Thermodynamic pressures divided by the QCD Stefan-Boltzmann (SB) limit are illustrated at $\mu=0 \mathrm{MeV}$ in Fig.\ref{Fig08} for three models, and the pressure has
already been normalized to vanish at $T=\mu=0$. For $N_f$ massless quarks and $N_c^2-1$ massless gluons in the deconfined phase, the QCD pressure in the SB limit is given by
\begin{eqnarray}
\frac{P_{\mathrm{SB}}}{T^4}=(N^2_c-1)\frac{\pi^2}{45}+N_cN_f\left[ \frac{7 \pi^2}{180}+\frac{1}{6}(\frac{\mu}{T})^2+\frac{1}{12\pi^2}(\frac{\mu}{T})^4\right],
\end{eqnarray}
where the first term is the gluonic contribution and the rest involves the fermions.

At very high temperature, e.g. around twice the chiral critical temperature, the pressure of the PQM tends to approach that of the NS model in the quark phase. This implies
that there only exists a weak interaction between particles in quark phase, and the quasi-particles model is a good approximation for the description of the weak-interaction
quark gluon plasma. However, when the temperature decreases, the $P_{\mathrm{PQM}}$ deviates from the pressure $P^{\mathrm{Q}}_{\mathrm{NS}}$ more and more. When the $T$
reaches out to the chiral critical temperature $T_{\chi}^c \sim 201 \mathrm{MeV}$, the gap between the two pressures arrives at its maximum value and then decrease smoothly
until zero temperature is reached. For comparing with lattice simulations with a temporal extent of $N_{\tau}=6$ which is closer to the continuum limit in
Ref.\cite{AliKhan:2001ek}, the strong interaction between the particles in the NS model cannot be simply discarded and it really plays a dominant role in producing a correct
and reasonable thermodynamical pressure of the system.

In Fig.\ref{Fig08}, we also plot the pressure as a function of the temperature starting from the hadron to the quark phase at $\mu_B=0$ for the NS model, while
varying the baryon masses for various temperature and densities in the confined phase. From the figure, the dash-dotted curve shows rapidly changed discontinuities at the
deconfinement critical temperature from hadron matter to quark matter. This indicates a pseudo first-order phase transition for the delocalization transition and signals a
drastic structural change for nucleon when the system goes with the diffusions of the solitons (nucleons) into thermal medium simultaneously. This strange behavior of the
pressure $P_{\mathrm{NS}}$ around the deconfinement critical temperature $T_{\chi}^c$ is believed to be removed by bringing in a self-consistent interaction of $\sigma$,
$\omega$, $\rho$ mesons with nuclear matter (solitons) in hadron phase. The strong interactions between mesons and nucleons, which are widely adopted in nuclear matter and
finite nuclei \cite{Celenza:1984ew,Jandel:1983gz,Wen:2008ui,Serot:1984ey,Guichon:1987jp,Saito:2005rv}, would provide necessary suppressions on the pressures for the mesons
and nucleons in hadron phase, and the remaining self-interaction of the mesons in quark phase could give further suppression on the pressure around $T_{\chi}^c$, similar to
the PQM model. Finally, it is worth to note that, besides the interactions, another important influence which was neglected here is the center-of mass (c.m.) correction to
the nucleon properties, which have been considered in soliton bag models and the QMC model largely\cite{Saito:2005rv,Kimel:1982ed,Lubeck:1986if,Lubeck:1987pj}. These two
corrections are certainly out of the scope of our current topic and we prefer leaving them for a future study.

Unlike the case in the quark phase, here the pressures of the PQM model and the nontopological soliton model are suppressed in the hadron phase and start to increase when
deconfinement sets in. The small difference among the pressures of the $P_{\mathrm{NS}}$ and $P_{\mathrm{PQM}}$ is due to the different treatment of the mesons in the hadron
phase. In the NS model the mesons are taken as the active degrees of freedom, but for the PQM model they are purely mean fields and should be restrained to the except values
when $T$ is close to the $T_{\chi}^c$. From Fig.\ref{Fig05}, it is shown that the effective nucleon mass $M_N$ slightly deviates from its vacuum value with temperature
increasing, only at the critical temperature $T^c_{\chi}$, $M_N$ experiences a sharp jump. Consequently, the contribution of nucleons to the total pressure in
Eq.(\ref{pressure}) is very small in hadron phase as far as the chemical potential is small. To estimate, it only gives $8.6\%$ contribution to $P_{\mathrm{NS}}$ when $T$ is
around $T_{d}^c\approx 177$ MeV for zero chemical potential. On the contrary, if three bound quarks are set to be free in quark phase, the valence quarks will give a dominant
contribution to the pressure in Eq.({\ref{pressure2}).

\section{Summary and discussion}
In the present paper we have investigated possible nontopological soliton solutions in  the effective potential of the PQM model in the presence of renormalized fermionic
vacuum. The results show that as long as the temperature is not larger than the chiral critical temperature $T_{\chi}^c$, there exist truly stable soliton solutions in the
model for both crossover and first-order phase transitions.

Even though there are stable soliton solutions for the Euler-Lagrange equations of the model at finite temperature and density, the stability of the solitons (nucleons) have
to be checked and analyzed carefully in thermal medium by comparing the effective masses of nucleons with the energies of three free constituent quarks. Our results show that
the chiral phase transition and the delocalization phase transition from nucleon matter to quark matter take place simultaneously for the first-order phase transition. For
$T<T_{\chi}^c$, the free constituent quarks are not the ground state of the strongly interacting system, and the quarks will reorganize so to form lower-energy bound states
carrying the hedgehog configuration. However, as soon as the temperature $T$ crosses over the $T_{\chi}^c$, such bound states cannot survive anymore, and the system
experiences a first-order hadron-quark phase transition to the chirally symmetric phase.

The situation differs from the case of the first-order transition, in the crossover transition, even though the effective potential genuinely ensures the stable soliton
solution in the system, but it is energetically unfavorable for nucleons to exist when $T$ is across the deconfinement critical temperature $T_{d}^c$. The difference between
$T_{\chi}^c$ and $T_{d}^c$ is about $24\mathrm{MeV}$ for zero chemical potential, but it will decrease to zero as $\mu$ increases to some value around $302 \mathrm{MeV}$. This
particular point sometimes is denoted as the critical end point (CEP), which appears as well in the phase diagram of Fig.\ref{Fig06} .

In order to compare our results with the lattice QCD simulations and other models at zero chemical potential but finite temperature directly, we have investigated the
thermodynamic properties of the nontopological soliton model in the PQM model. It is found that the inclusion of the Polyakov loop is necessary and important when comparing
with the lattice QCD simulations. When compared with the previous studies in this field, we notice a quite improvement on the topic by providing a reasonable critical
deconfinement temperature $T_d^c\sim 177 \mathrm{MeV}$ for $\mu=0$ and extracting a standard QCD phase diagram in agreement with the lattice data and other phenomenological
models' predictions\cite{Fukushima:2010bq}. However, the strange behavior of the performed pressure in the hadron-quark phase transition indicates that the description of the
hadron phase as a non-interacting hadron gas of the nucleons and mesons with medium-modified masses has underestimated the important effects of their interactions, and these
interactions should be introduced to further suppress the pressures of the mesons and nucleons both in hadron and quark phase. In other words, the present form of study is a
prototype and still not suitable for the proper description of nuclear matter and finite nuclei in hadron phase yet, and it deserves further efforts on making the model
applicable for hadron-quark phase transition completely and satisfactorily. Eventually, work in this direction is in progress.

\begin{acknowledgments}
We thank Jingniu Hu for valuable comments and discussions. Authors are grateful to Prof. Yijin Yan for his kind
hospitality during our visiting in Hong Kong University of Science and Technology. The project is Supported in part by NSFC under No.11274085 and 11275002.
\end{acknowledgments}


\begin{thebibliography}{199}

  %\cite{Rischke:2003mt}
\bibitem{Rischke:2003mt}
  D.~H.~Rischke,
%  ``The quark-gluon plasma in equilibrium,''
  Prog.\ Part.\ Nucl.\ Phys.\  {\bf 52}, 197 (2004).

%\cite{Yagi:2005yb}
\bibitem{Yagi:2005yb}
  K.~Yagi, T.~Hatsuda and Y.~Miake,
  ``\textit{Quark-gluon plasma: From big bang to little bang},''
  Camb.\ Monogr.\ Part.\ Phys.\ Nucl.\ Phys.\ Cosmol.\  {\bf 23}, 1 (2005).  %%CITATION = CMPCE,23,1;%%

%\cite{Fukushima:2010bq}
\bibitem{Fukushima:2010bq}
  K.~Fukushima and T.~Hatsuda,
  %``The phase diagram of dense QCD,''
  Rept.\ Prog.\ Phys.\  {\bf 74}, 014001 (2011).  %%CITATION = ARXIV:1005.4814;%%

  %\cite{Chodos:1974je}
\bibitem{Chodos:1974je}
  A.~Chodos, R.~L.~Jaffe, K.~Johnson, C.~B.~Thorn and V.~F.~Weisskopf,
  %``A New Extended Model of Hadrons,''
  Phys.\ Rev.\ D {\bf 9}, 3471 (1974);
    %\cite{Chodos:1974pn}
%\bibitem{Chodos:1974pn}
  A.~Chodos, R.~L.~Jaffe, K.~Johnson and C.~B.~Thorn,
  %``Baryon Structure in the Bag Theory,''
  Phys.\ Rev.\ D {\bf 10}, 2599 (1974);
  %\cite{DeGrand:1975cf}
%\bibitem{DeGrand:1975cf}
  T.~A.~DeGrand, R.~L.~Jaffe, K.~Johnson and J.~E.~Kiskis,
  %``Masses and Other Parameters of the Light Hadrons,''
  Phys.\ Rev.\ D {\bf 12}, 2060 (1975).

  %\cite{Nambu:1961tp}
\bibitem{Nambu:1961tp}
  Y.~Nambu and G.~Jona-Lasinio,
  %``Dynamical Model of Elementary Particles Based on an Analogy with Superconductivity. 1.,''
  Phys.\ Rev.\  {\bf 122}, 345 (1961);
    %\cite{Nambu:1961fr}
%\bibitem{Nambu:1961fr}
  Y.~Nambu and G.~Jona-Lasinio,
  %``Dynamical Model Of Elementary Particles Based On An Analogy With Superconductivity. Ii,''
  Phys.\ Rev.\  {\bf 124}, 246 (1961).

  %\cite{Vogl:1991qt}
\bibitem{Vogl:1991qt}
  U.~Vogl and W.~Weise,
  %``The Nambu and Jona Lasinio model: Its implications for hadrons and nuclei,''
  Prog.\ Part.\ Nucl.\ Phys.\  {\bf 27}, 195 (1991);  %%CITATION = PPNPD,27,195;%%
%\cite{Klevansky:1992qe}
%\bibitem{Klevansky:1992qe}
  S.~P.~Klevansky,
  %``The Nambu-Jona-Lasinio model of quantum chromodynamics,''
  Rev.\ Mod.\ Phys.\  {\bf 64}, 649 (1992);  %%CITATION = RMPHA,64,649;%%
%\cite{Hatsuda:1994pi}
%\cite{Hatsuda:1994pi}
%\bibitem{Hatsuda:1994pi}
  T.~Hatsuda and T.~Kunihiro,
  %``QCD phenomenology based on a chiral effective Lagrangian,''
  Phys.\ Rept.\  {\bf 247}, 221 (1994);  %%CITATION = HEP-PH/9401310;%%
%\cite{Buballa:2003qv}
%\bibitem{Buballa:2003qv}
  M.~Buballa,
  %``NJL model analysis of quark matter at large density,''
  Phys.\ Rept.\  {\bf 407}, 205 (2005).  %%CITATION = HEP-PH/0402234;%%

 %\cite{GellMann:1960np}
\bibitem{GellMann:1960np}
  M.~Gell-Mann and MLevy,
  %``The axial vector current in beta decay,''
  Nuovo Cim.\  {\bf 16}, 705 (1960).  %%CITATION = NUCIA,16,705;%%

  %\cite{Ring:1980}
\bibitem{Ring:1980}
 P. Ring and P. Schuck, ``\textit{The Nuclear Many-Body Problem},''
(Springer, Heidelberg, 1980).

  %\cite{Serot:1986}
\bibitem{Serot:1986}
B. D. Serot and J. D. Walecka, ``\textit{Advances in Nuclear Physics},'' (Plenum, New York, 1986), Vol. 16.

%\cite{Yasutake:2012dw}
\bibitem{Yasutake:2012dw}
  N.~Yasutake, T.~Noda, H.~Sotani, T.~Maruyama and T.~Tatsumi,
  %``Thermodynamical description of hadron-quark phase transition and its implications on compact-star phenomena,''
  arXiv:1208.0427 [astro-ph.HE].

  %\cite{Lourenco:2012dx}
\bibitem{Lourenco:2012dx}
  O.~Lourenco, M.~Dutra, A.~Delfino and M.~Malheiro,
  %``Hadron-quark phase transition in a hadronic and Polyakov--Nambu--Jona-Lasinio models perspective,''
  Phys.\ Rev.\ D {\bf 84}, 125034 (2011).  %%CITATION = ARXIV:1201.1239;%%

  %\cite{Shao:2011fk}
\bibitem{Shao:2011fk}
  G.~Y.~Shao, M.~Di Toro, V.~Greco, M.~Colonna, S.~Plumari, B.~Liu and Y.~X.~Liu,
  %``Phase diagrams in the Hadron-PNJL model,''
  Phys.\ Rev.\ D {\bf 84}, 034028 (2011).

 %\cite{Costa:2010zw}
\bibitem{Costa:2010zw}
  P.~Costa, M.~C.~Ruivo, C.~A.~de Sousa and H.~Hansen,
  %``Phase diagram and critical properties within an effective model of QCD: the Nambu-Jona-Lasinio model coupled to the Polyakov loop,''
  Symmetry {\bf 2}, 1338 (2010), and references therein.

  %\cite{Herbst:2013ail}
\bibitem{Herbst:2013ail}
  T.~K.~Herbst, J.~M.~Pawlowski and B.~J.~Schaefer,
  %``Phase structure and thermodynamics of QCD,''
  Phys.\ Rev.\ D {\bf 88}, no. 1, 014007 (2013).

  %\cite{Schaefer:2007pw}
\bibitem{Schaefer:2007pw}
  B.~J.~Schaefer, J.~M.~Pawlowski and J.~Wambach,
  %``The Phase Structure of the Polyakov--Quark-Meson Model,''
  Phys.\ Rev.\  D {\bf 76}, 074023 (2007).

  %\cite{Marko:2010cd}
\bibitem{Marko:2010cd}
  G.~Marko and Z.~Szep,
  %``Influence of the Polyakov loop on the chiral phase transition in the two flavor chiral quark model,''
  Phys.\ Rev.\ D {\bf 82}, 065021 (2010).

 %\cite{Herbst:2010rf}
\bibitem{Herbst:2010rf}
  T.~K.~Herbst, J.~M.~Pawlowski and B.~J.~Schaefer,
  %``The phase structure of the Polyakov--quark-meson model beyond mean field,''
  Phys.\ Lett.\ B {\bf 696}, 58 (2011).

 %\cite{Gupta:2011ez}
\bibitem{Gupta:2011ez}
  U.~S.~Gupta and V.~K.~Tiwari,
  %``Revisiting the Phase Structure of the Polyakov-quark-meson Model in the presence of Vacuum Fermion Fluctuation,''
  Phys.\ Rev.\ D {\bf 85}, 014010 (2012).
  %%CITATION = ARXIV:1107.1312;%%

  %\cite{Skokov:2010sf}
\bibitem{Skokov:2010sf}
  V.~Skokov, B.~Friman, E.~Nakano, K.~Redlich and B.-J.~Schaefer,
  %``Vacuum fluctuations and the thermodynamics of chiral models,''
  Phys.\ Rev.\ D {\bf 82}, 034029 (2010).
  %%CITATION = ARXIV:1005.3166;%%

  %\cite{Tiwari:2012yy}
\bibitem{Tiwari:2012yy}
  V.~K.~Tiwari,
  %``Exploring criticality in the QCD-like two quark flavour models,''
  Phys.\ Rev.\ D {\bf 86}, 094032 (2012).
  %%CITATION = ARXIV:1208.2458;%%

  %\cite{Mao:2009aq}
\bibitem{Mao:2009aq}
  H.~Mao, J.~Jin and M.~Huang,
  %``Phase diagram and thermodynamics of the Polyakov linear sigma model with three quark flavors,''
  J.\ Phys.\ G {\bf 37}, 035001 (2010).  %%CITATION = ARXIV:0906.1324;%%

  %\cite{Gupta:2009fg}
\bibitem{Gupta:2009fg}
  U.~S.~Gupta and V.~K.~Tiwari,
  %``Meson Masses and Mixing Angles in 2+1 Flavor Polyakov Quark Meson Sigma Model and Symmetry Restoration Effects,''
  Phys.\ Rev.\ D {\bf 81}, 054019 (2010).
  %%CITATION = ARXIV:0911.2464;%%

  %\cite{Schaefer:2009ui}
\bibitem{Schaefer:2009ui}
  B.~J.~Schaefer, M.~Wagner and J.~Wambach,
  %``Thermodynamics of (2+1)-flavor QCD: Confronting Models with Lattice Studies,''
  Phys.\ Rev.\ D {\bf 81}, 074013 (2010).
  %%CITATION = ARXIV:0910.5628;%%

  %\cite{Birse:1983gm}
\bibitem{Birse:1983gm}
  M.~C.~Birse and M.~K.~Banerjee,
  %``A Chiral Soliton Model of Nucleon and Delta,''
  Phys.\ Lett.\ B {\bf 136}, 284 (1984);  %%CITATION = PHLTA,B136,284;%%
%\cite{Birse:1984js}
%\bibitem{Birse:1984js}
  M.~C.~Birse and M.~K.~Banerjee,
  %``A Chiral Model for Nucleon and Delta,''
  Phys.\ Rev.\ D {\bf 31}, 118 (1985).

  %\cite{Kahana:1984dx}
\bibitem{Kahana:1984dx}
  S.~Kahana, G.~Ripka and V.~Soni,
  %``Soliton with Valence Quarks in the Chiral Invariant Sigma Model,''
  Nucl.\ Phys.\ A {\bf 415}, 351 (1984).

 %\cite{Alberto:1988xj}
\bibitem{Alberto:1988xj}
  P.~Alberto, E.~Ruiz Arriola, M.~Fiolhais, F.~Grummer, J.~N.~Urbano and K.~Goeke,
  %``Nucleon Form-factors In The Projected Linear Chiral Soliton Model,''
  Phys.\ Lett.\ B {\bf 208}, 75 (1988).  %%CITATION = PHLTA,B208,75;%%

%\cite{Bernard:1988db}
\bibitem{Bernard:1988db}
  V.~Bernard and U.~G.~Meissner,
  %``Properties of Vector and Axial Vector Mesons from a Generalized Nambu-Jona-Lasinio Model,''
  Nucl.\ Phys.\ A {\bf 489}, 647 (1988).  %%CITATION = NUPHA,A489,647;%%

%\cite{Alberto:1990ru}
\bibitem{Alberto:1990ru}
  P.~Alberto, E.~Ruiz Arriola, M.~Fiolhais, K.~Goeke, F.~Grummer and J.~N.~Urbano,
  %``Form-factors in the projected linear chiral sigma model,''
  Z.\ Phys.\ A {\bf 336}, 449 (1990).  %%CITATION = ZEPYA,A336,449;%%

%\cite{Goeke:1988hp}
\bibitem{Goeke:1988hp}
  K.~Goeke, M.~Harvey, F.~Grummer and J.~N.~Urbano,
  %``Chiral Model Of The Nucleon And Delta Isobar: The Coherent Pair Approximation,''
  Phys.\ Rev.\ D {\bf 37}, 754 (1988).  %%CITATION = PHRVA,D37,754;%%

%\cite{Aly:1998wg}
\bibitem{Aly:1998wg}
  T.~S.~T.~Aly, J.~A.~McNeil and S.~Pruess,
  %``Chiral baryon in the coherent pair approximation,''
  Phys.\ Rev.\ D {\bf 60}, 114022 (1999).  %%CITATION = HEP-PH/9809473;%%

  %\cite{Christov:1991ry}
\bibitem{Christov:1991ry}
  C.~V.~Christov, E.~Ruiz Arriola and K.~Goeke,
  %``Nucleon properties and restoration of chiral symmetry at finite density and temperature in Nambu-Jona-Lasinio model,''
  Nucl.\ Phys.\ A {\bf 556}, 641 (1993).  %%CITATION = HEP-PH/9303213;%%

%\cite{AbuShady:2012zza}
\bibitem{AbuShady:2012zza}
  M.~Abu-Shady and H.~M.~Mansour,
  %``Quantized linear sigma model at finite temperature, and nucleon properties,''
  Phys.\ Rev.\ C {\bf 85}, 055204 (2012).  %%CITATION = PHRVA,C85,055204;%%

    %\cite{Hong1997clj}
\bibitem{Hong1997clj}
  H. Chen, B. Liu and H. Jiang
%``Light Hadrons at Finite Temperature in the Linear Sigma Model,''
  Chin. Phys. Lett. \textbf{14}, 645 (1997).

  %\cite{Mao:2013qu}
\bibitem{Mao:2013qu}
  H.~Mao, T.~Wei and J.~Jin,
  %``Chiral soliton model at finite temperature and density,''
  Phys.\ Rev.\ C {\bf 88}, 035201 (2013).

  %\cite{Zhang:2015vva}
\bibitem{Zhang:2015vva}
  H.~Zhang, R.~Dong and S.~Shu,
  %``The baryon mass calculation in the chiral soliton model at finite temperature and density,''
  Int.\ J.\ Mod.\ Phys.\ E {\bf 24}, 1550025 (2015).

  %\cite{Mansour:2015yha}
\bibitem{Mansour:2015yha}
  H.~M.~Mansour and M.~Abu-Shady,
  %``Nucleon Properties in the Quantized Linear Sigma Model at Finite Temperature and Chemical Potential,''
  arXiv:1507.02214 [hep-ph].

  %\cite{Reinhardt:1985nq}
\bibitem{Reinhardt:1985nq}
  H.~Reinhardt, B.~V.~Dang and H.~Schulz,
  %``Deconfinement Phase Transition Of Hot And Dense Nuclear Matter In The
  %Nontopological Soliton Bag Model,''
  Phys.\ Lett.\  B {\bf 159}, 161 (1985);
%\cite{Mao:2007gm}
%\bibitem{Mao:2007gm}
  H.~Mao, M.~Yao and W.~-Q.~Zhao,
  %``The Friedberg-Lee model at finite temperature and density,''
  Phys.\ Rev.\ C {\bf 77}, 065205 (2008).  %%CITATION = ARXIV:0711.4643;%%

%\cite{Gao:1992zd}
\bibitem{Gao:1992zd}
  S. Gao, E. Wang and J. Li,
  %``Bag constant and deconfinement phase transition in a nontopological soliton
  %model,''
  Phys.\ Rev.\ D {\bf 46}, 3211 (1992).

%\cite{Li:1987wb}
\bibitem{Li:1987wb}
  M.~Li, M.~C.~Birse and L.~Wilets,
  %``Phase Transition In The Soliton Bag Model,''
  J.\ Phys.\ G {\bf 13} (1987) 1.

 %\cite{Mao:2006zp}
\bibitem{Mao:2006zp}
  H.~Mao, R.~-K.~Su and W.~-Q.~Zhao,
  %``Soliton solutions of the improved quark mass density-dependent model at finite temperature,''
  Phys.\ Rev.\ C {\bf 74}, 055204 (2006).  %%CITATION = HEP-PH/0606239;%%
  %%%%%%%%%%%%%%%%%%%%%%%%%%%%%%%%%%%%%%%%

  %\cite{Fukushima:2003fw}
\bibitem{Fukushima:2003fw}
  K.~Fukushima,
  %``Chiral effective model with the Polyakov loop,''
  Phys.\ Lett.\  B {\bf 591}, 277 (2004).

  %\cite{Polyakov:1978vu}
\bibitem{Polyakov:1978vu}
  A.~M.~Polyakov,
  %``Thermal Properties Of Gauge Fields And Quark Liberation,''
  Phys.\ Lett.\  B {\bf 72} (1978) 477.

%\cite{Ratti:2005jh}
\bibitem{Ratti:2005jh}
  C.~Ratti, M.~A.~Thaler and W.~Weise,
  %``Phases of QCD: Lattice thermodynamics and a field theoretical model,''
  Phys.\ Rev.\  D {\bf 73}, 014019 (2006).

%\cite{Scavenius:2000qd}
\bibitem{Scavenius:2000qd}
  O.~Scavenius, A.~Mocsy, I.~N.~Mishustin and D.~H.~Rischke,
  %``Chiral phase transition within effective models with constituent quarks,''
  Phys.\ Rev.\ C {\bf 64}, 045202 (2001).  %%CITATION = NUCL-TH/0007030;%%

  %\cite{Kapusta:2006pm}
\bibitem{Kapusta:2006pm}
  J.~I.~Kapusta and C.~Gale,
  ``\textit{Finite-temperature field theory: Principles and applications},''
%\href{http://www.slac.stanford.edu/spires/find/hep/www?irn=7209002}{SPIRES entry}
( Cambridge University Press, UK, 2006).

%\cite{Agashe:2014kda}
\bibitem{Agashe:2014kda}
  K.~A.~Olive {\it et al.} [Particle Data Group Collaboration],
  %``Review of Particle Physics,''
  Chin.\ Phys.\ C {\bf 38}, 090001 (2014).
  %%CITATION = CHPHD,C38,090001;%%

  %\cite{Fukushima:2008wg}
\bibitem{Fukushima:2008wg}
  K.~Fukushima,
  %``Phase diagrams in the three-flavor Nambu--Jona-Lasinio model with the
  %Polyakov loop,''
  Phys.\ Rev.\  D {\bf 77}, 114028 (2008);
  [Erratum-ibid.\  D {\bf 78}, 039902 (2008)].

  %\cite{Friedberg:1976eg}
\bibitem{Friedberg:1976eg}
  R.~Friedberg and T.~D.~Lee,
  %``Fermion Field Nontopological Solitons. 1,''
  Phys.\ Rev.\ D {\bf 15}, 1694 (1977); R.~Friedberg and T.~D.~Lee,
  %``Fermion Field Nontopological Solitons. 2. Models For Hadrons,''
  Phys.\ Rev.\ D {\bf 16}, 1096 (1977); R.~Friedberg and T.~D.~Lee,
  %``QCD And The Soliton Model Of Hadrons,''
  Phys.\ Rev.\ D {\bf 18}, 2623 (1978).



  %\cite{Birse:1991cx}
\bibitem{Birse:1991cx}
  M.~C.~Birse,
  %``Soliton models for nuclear physics,''
  Prog.\ Part.\ Nucl.\ Phys.\  {\bf 25}, 1 (1990).  %%CITATION = PPNPD,25,1;%%

  %\cite{Goldflam:1981tg}
\bibitem{Goldflam:1981tg}
  R.~Goldflam and L.~Wilets,
  %``The Soliton Bag Model,''
  Phys.\ Rev.\ D {\bf 25}, 1951 (1982).
  %%CITATION = PHRVA,D25,1951;%%
  %189 citations counted in INSPIRE as of 29 Jul 2015

  %\cite{Berger:1996hc}
\bibitem{Berger:1996hc}
  J.~Berger and C.~V.~Christov,
  %``Phase transition and nucleon as soliton in the Nambu-Jona-Lasinio model at finite temperature and density,''
  Nucl.\ Phys.\ A {\bf 609}, 537 (1996).  %%CITATION = HEP-PH/9607219;%%

%\cite{Schleif:1997pi}
\bibitem{Schleif:1997pi}
  M.~Schleif and R.~Wunsch,
  %``Thermodynamic properties of the SU(2)(f) chiral quark loop soliton,''
  Eur.\ Phys.\ J.\ A {\bf 1}, 171 (1998).

 %\cite{Ascher:1981}
\bibitem{Ascher:1981}
 U. Ascher, J. Christiansen, R.D. Russell, R.D,
 %``Collocation software for boundary-value odes,''
ACM Trans. Math. Softw. {\bf{7}}(2), 209–222 (1981).

%\cite{Gorenstein:1995vm}
\bibitem{Gorenstein:1995vm}
  M.~I.~Gorenstein and S.~-N.~Yang,
  %``Gluon plasma with a medium dependent dispersion relation,''
  Phys.\ Rev.\ D {\bf 52}, 5206 (1995).

  %\cite{AliKhan:2001ek}
\bibitem{AliKhan:2001ek}
  A.~Ali Khan {\it et al.} [CP-PACS Collaboration],
  %``Equation of state in finite temperature QCD with two flavors of improved Wilson quarks,''
  Phys.\ Rev.\ D {\bf 64}, 074510 (2001).

%\cite{Celenza:1984ew}
\bibitem{Celenza:1984ew}
  L.~S.~Celenza, A.~Rosenthal and C.~M.~Shakin,
  %``Symmetry Breaking, Quark Deconfinement And The Emc Effect,''
Phys.\ Rev.\ Lett.\  {\bf 53}, 892 (1984).  %%CITATION = PRLTA,53,892;%%

%\cite{Jandel:1983gz}
\bibitem{Jandel:1983gz}
  M.~Jandel and G.~Peters,
  %``The Nuclear Quark Confinement Size In A Soliton Bag Model,''
  Phys.\ Rev.\ D {\bf 30}, 1117 (1984).  %%CITATION = PHRVA,D30,1117;%%

%\cite{Wen:2008ui}
\bibitem{Wen:2008ui}
  W.~Wen and H.~Shen,
  %``Modification of nucleon properties in nuclear matter and finite nuclei,''
  Phys.\ Rev.\ C {\bf 77}, 065204 (2008).  %%CITATION = ARXIV:0806.0504;%%

    %\cite{Serot:1984ey}
\bibitem{Serot:1984ey}
  B.~D.~Serot and J.~D.~Walecka,
  %``The Relativistic Nuclear Many Body Problem,''
  Adv.\ Nucl.\ Phys.\  {\bf 16}, 1 (1986).  %%CITATION = ANUPB,16,1;%%

%\cite{Guichon:1987jp}
\bibitem{Guichon:1987jp}
  P.~A.~M.~Guichon,
  %``A Possible Quark Mechanism for the Saturation of Nuclear Matter,''
  Phys.\ Lett.\ B {\bf 200}, 235 (1988).  %%CITATION = PHLTA,B200,235;%%

%\cite{Saito:2005rv}
\bibitem{Saito:2005rv}
  K.~Saito, K.~Tsushima and A.~W.~Thomas,
  %``Nucleon and hadron structure changes in the nuclear medium and impact  on
  %observables,''
  Prog.\ Part.\ Nucl.\ Phys.\  {\bf 58}, 1 (2007).

  %\cite{Kimel:1982ed}
\bibitem{Kimel:1982ed}
  I.~Kimel,
  %``New Formula For Proton - Neutron Mass Difference,''
  Phys.\ Rev.\ D {\bf 27}, 2129 (1983).

  %\cite{Lubeck:1986if}
\bibitem{Lubeck:1986if}
  E.~G.~Lubeck, M.~C.~Birse, E.~M.~Henley and L.~Wilets,
  %``Momentum Projection and Relativistic Boost of Solitons: Coherent States and Projection,''
  Phys.\ Rev.\ D {\bf 33}, 234 (1986).

  %\cite{Lubeck:1987pj}
\bibitem{Lubeck:1987pj}
  E.~G.~Lubeck, E.~M.~Henley and L.~Wilets,
  %``Momentum Projection and Relativistic Boost of Solitons. 2. Nucleon and Meson Properties,''
  Phys.\ Rev.\ D {\bf 35}, 2809 (1987).


\end{thebibliography}
\end{document}